\documentclass[12pt]{article}

\usepackage{newtxtext,newtxmath}
\usepackage{nicefrac}
\usepackage{soul}

\usepackage{graphicx}

\usepackage[letterpaper,margin=1in]{geometry}

\renewenvironment{abstract}
	{\quotation}
	{\endquotation}

\date{}

\makeatletter
\renewcommand{\fnum@figure}{\textbf{Figure \thefigure}}
\renewcommand{\fnum@table}{\textbf{Table \thetable}}
\makeatother

\usepackage{scicite}

\usepackage{url}

\usepackage{caption}
\DeclareCaptionFormat{breakable}{#1#2#3\par}
\renewcommand{\vec}[1]{\mathbf{#1}}

\def\scititle{
    Hierarchical quasiparticle dynamics in antiferromagnets revealed by time- and momentum-resolved X-ray scattering
}
\title{\bfseries \boldmath \scititle}

\author{
	Arnau~Romaguera$^{1}$, 
	Elizabeth~Skoropata$^{1}$,
    Yun~Yen$^{2,3,4}$,
    Biaolong~Liu$^{1}$,\and
    Abhishek~Nag$^{1, \dagger}$,
    Shih-Wen~Huang$^{1}$,
    Ludmila~Leroy$^{1}$,
    Katja~Sophia~Moos$^{2, 5}$,
    Gian~Parusa$^{2,5}$,\and
    Serhane~Zerdane$^{1}$,
    Ritwika~Mandal$^{6}$,
    Céline~Mariette$^{7}$,
    Matteo~Levantino$^{7}$,\and
    Eugenio~Paris$^{1}$, 
    Luc~Patthey$^{1}$,
    Ekaterina~Pomjakushina$^{8}$,
	Urs~Staub$^{1}$,\and
    Monica~Ciomaga~Hatnean$^{8,9}$,
    Michael~Schüler$^{2,5}$,
    Elia~Razzoli$^{1}$,
    and Hiroki~Ueda$^{1\ast}$\and
	\small$^{1}$PSI Center for Photon Science, Paul Scherrer Institute, 5232 Villigen-PSI, Switzerland.\and
	\small$^{2}$PSI Center for Scientific Computing, Theory and Data, Paul Scherrer Institute, 5232 Villigen-PSI, Switzerland.\and
    \small$^{3}$École Polytechnique Fédérale de Lausanne (EPFL), 1015 Lausanne, Switzerland.\and
    \small$^{4}$Institute for Theoretical Physics and Bremen Center for Computational Materials Science, \\
    \small University of Bremen, 28359 Bremen, Germany.\and
    \small$^{5}$Department of Physics, University of Fribourg, 1700 Fribourg, Switzerland.\and
    \small$^{6}$Institut des Matériaux de Nantes Jean Rouxel (IMN) – UMR 6502, Nantes Université, .\and
    \small CNRS, 44000 Nantes, France.\and
    \small$^{7}$ESRF – The European Synchrotron, 71 Avenue des Martyrs, CS40220, 38043, Grenoble, Cedex, France.\and
    \small$^{8}$PSI Center for Neutron and Muon Sciences, Paul Scherrer Institute, 5232 Villigen-PSI, Switzerland.\and
    \small$^{9}$Materials Discovery Laboratory, Department of Materials, ETH Zürich, 8093 Zürich, Switzerland.\and
	\small$^\ast$Corresponding author. Email: hiroki.ueda@psi.ch\and
	\small$^\dagger$Present address: Department of Physics, Indian Institute of Technology, Roorkee, Uttarakhand 247667, India.
}

\begin{document} 

% Insert the title and author list
\maketitle

% Abstract, in bold
% There are strict length limits, and not all formats have abstracts.
% Consult the journal instructions to authors for details.
% Do not cite any references in the abstract.
\begin{abstract} \bfseries \boldmath
% % Start with one or two sentences of background
% This is a simple template to prepare papers in \LaTeX\ for the \textit{Science}-family journals.
% Abstracts start with one or two sentences of background, which should be
% comprehensible to any scientist.
% % Then summarise the results of your observations, experiments, simulations etc.
% The following text should outline the main results of the research.
% Simple mathematical expressions can be included e.g. $a^2+b^2=c^2$.
% % End with a statement of your main conclusions
% The final sentence of the abstract should state the main conclusions and implications.
Energy flows among coupled subsystems are essential for ultrafast dynamics and high-speed technologies. In magnetic materials, spin fluctuations ---magnons--- mediate these flows in ultrafast magnetism. Yet momentum-resolved access to low-energy magnons governing the microscopic dynamics has been lacking. Using time-resolved resonant diffuse scattering alongside complementary time-resolved X-ray techniques and quantum-kinetic simulations, we unveil the hierarchical energy pathways among correlated systems in the photoexcited antiferromagnet CuO. Above-bandgap excitation triggers near-instantaneous spin disorder, generating non-thermal magnons throughout reciprocal space within femtoseconds. Real-time momentum-resolved tracking reveals picosecond magnon quasi-thermalization, followed by nanosecond recovery via momentum-selective magnon-phonon scattering. The quasiparticle dispersion mismatch creates recovery bottlenecks that control non-equilibrium lifetimes. This microscopic framework transcends phenomenological models and generalizes across materials, establishing design principles for ultrafast control of material properties.
\end{abstract}

\noindent
Ultrafast manipulation of material states offers a powerful pathway to tailor material properties by driving systems into non-equilibrium states that are otherwise inaccessible~\cite{Li1, Mitrano1, Li2, Kogar1, Disa1, Ilyas1, Zeng1}. Understanding the microscopic mechanisms underlying these ultrafast phenomena is essential for fundamental physics and next-generation high-speed technologies. Ultrafast magnetism has emerged as a particularly compelling research area, encompassing phenomena such as ultrafast demagnetization~\cite{Beaurepaire1}, coherent magnon creation~\cite{Kimel1}, photo-induced phase transitions~\cite{Li1}, spin current generation~\cite{Kampfrath1}, all-optical magnetization switching~\cite{Lambert1}, and ultrafast modifications of spin fluctuations~\cite{Padma1}, which can influence emergent phenomena, including superconductivity~\cite{Fong1}, Peierls transitions~\cite{Hase1}, and transport~\cite{Desjardins1}. This breadth of phenomena highlights the significant potential for technological applications in spintronics, magnonics, and ultrafast magnetic storage, electronics, and logic devices. 

Ultrafast magnetism typically involves complex interactions among multiple subsystems --spins, charges, orbitals, and the lattice-- rather than occurring solely within the spin sector. While the multi-temperature model provides a qualitative framework for describing these coupled dynamics~\cite{Beaurepaire1}, detailed microscopic insights require understanding the specific roles of other degrees of freedom, including phonons~\cite{Afanasiev1, Tauchert1, Ueda1}, magnetic anisotropy~\cite{Stupakiewicz1}, and charge dynamics~\cite{Kim1}. The intricate interactions following photoexcitation clarify leading mechanisms in fundamental magnetism, yet a comprehensive understanding demands tracking the dynamics of these interconnected subsystems and their hierarchical energy transfer pathways. 

Here, we selectively investigate the transient behavior of the electronic, magnetic, and lattice systems associated with ultrafast sublattice demagnetization in the antiferromagnetic (AFM) insulator cupric oxide (CuO). We combine three complementary time-resolved X-ray techniques, each with specific sensitivity to different degrees of freedom, enabling us to disentangle hierarchical interactions and track energy flow pathways. Together with microscopic quantum-kinetic simulations (photoexcited Hubbard model and quantum Boltzmann equation) and density-functional theory (DFT) calculations, these measurements establish a generalizable, comprehensive microscopic framework for the excitation and recovery processes of ultrafast sublattice demagnetization in AFM insulators. 

\begin{figure}
	\centering
	\includegraphics[width=0.9\textwidth]{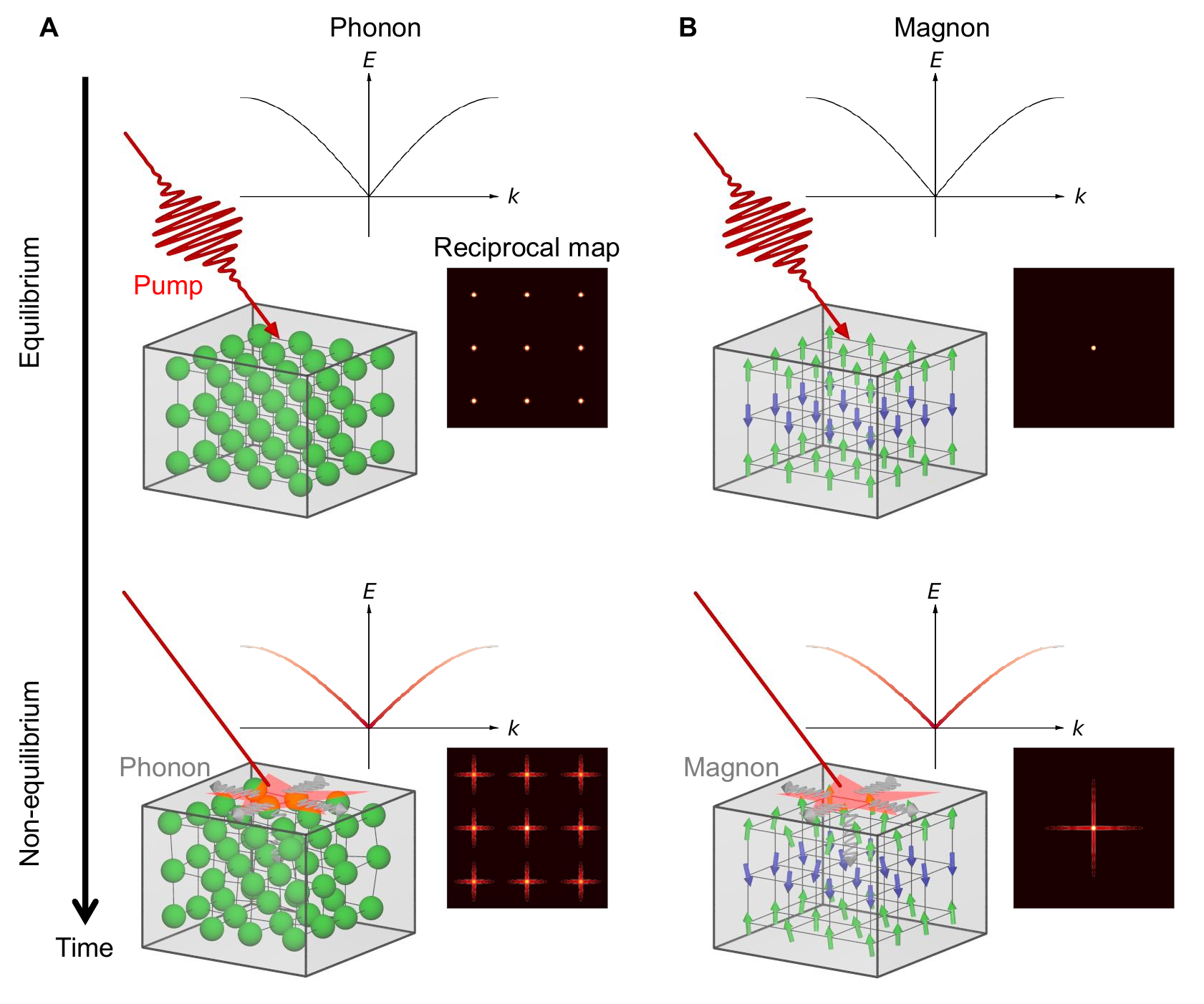} 
	% Pick an appropriate width - in print, figures are usually one or two columns wide, which can
	% be approximated by 0.3\textwidth or 0.6\textwidth respectively. Use appropriate label sizes.
	\caption{\textbf{Schematic illustration of time-resolved diffuse scattering.}
		(\textbf{A}) Time-resolved non-resonant diffuse scattering to probe phonons. (\textbf{B}) Time-resolved resonant diffuse scattering to probe magnons. (Top) Periodic arrangement of atoms or spins produces corresponding Bragg reflections in reciprocal space and low-energy collective excitations. (Bottom) Photoexcitation shakes the atomic positions or randomizes spin orientations, represented by the creation of low-energy quasiparticles along the dispersion curve, i.e., phonons and magnons, respectively. The enhanced density of low-energy quasiparticles results in diffuse scattering around Bragg reflections as streaks, enabling momentum-resolved tracking of collective excitation dynamics in non-equilibrium states.}
	\label{fig:1}
\end{figure}

Time-resolved \textit{non-resonant} diffuse scattering (tr-NRDS) measures time- and momentum-dependent low-energy phonons, which appear as diffuse streaks around Bragg reflections~\cite{Trigo1, Monti1} (Fig.~\ref{fig:1}\textbf{A}) owing to the charge sensitivity of non-resonant X-ray scattering. In contrast, tuning the incident X-ray energy to an atomic resonance provides sensitivity to electronic degrees of freedom. Time-resolved resonant X-ray diffraction (tr-RXD) thus directly probes the time evolution of magnetic order~\cite{Johnson1, Kubacka1, Ueda1}. In analogy, time-resolved \textit{resonant} diffuse scattering (tr-RDS) is expected to access time- and momentum-resolved information on low-energy magnons (Fig.~\ref{fig:1}\textbf{B}), which would create diffuse streaks around magnetic Bragg reflections, though it has not yet been experimentally demonstrated. This emerging technique represents a powerful tool for revealing magnon dynamics in non-equilibrium states, providing unprecedented insights into ultrafast spin dynamics.

\subsection*{Resonant diffuse scattering of non-equilibrium magnon origin}

Monoclinic CuO [space group $C2/c$~\cite{Forsyth1}] exhibits collinear AFM order (Fig.~\ref{fig:2}\textbf{A}), manifested by the $(\nicefrac{1}{2}\ 0\ {-\nicefrac{1}{2}})$ magnetic Bragg reflection at the Cu $L_3$ edge~\cite{Scagnoli1}. The electronic structure indicates that CuO is a charge-transfer insulator~\cite{Ghijsen1}, as consistent with our DFT calculations (Fig.~\ref{fig:band_structure}). A femtosecond laser pulse with a wavelength of 400 nm photo-dopes electrons into the upper Hubbard band of Cu $3d$ orbitals via charge transfer excitation from O $2p$ orbitals (Fig.~\ref{fig:2}\textbf{B}). The resultant doublons (doubly occupied electrons on Cu sites) quench exchange interactions via spin-charge coupling~\cite{Werner1, Mentink1} (Supplementary Text) and trigger ultrafast sublattice demagnetization on a femtosecond timescale set by the electron hopping during the pump duration. Indeed, the magnetic Bragg reflection intensity quasi-instantaneously decreases in a step-function-like manner within $\sim$70 fs (Fig.~\ref{fig:2}\textbf{D}), consistent with the previous observation with an 800 nm excitation~\cite{Johnson1}. Typical doublon lifetimes in copper-based oxides are within $\sim$200 fs~\cite{Novelli1, Baykusheva1}. 
Note that this ultrafast sublattice demagnetization process, driven by the intrinsic spin-charge coupling in Mott insulators, differs fundamentally from phonon-mediated mechanisms in itinerant ferromagnets~\cite{Sharma1, Weißenhofer1}.

Remarkably, the resonant X-ray scattering intensity measured simultaneously at $(0.49, 0.01, -0.51)$, slightly detuned from the Bragg point (Fig.~\ref{fig:S1}), shows a substantial upturn signal above the base line after an initial fast decrease (Fig.~\ref{fig:2}\textbf{D}), which is due to the magnetic Bragg reflection tail. Fitting to two exponential functions (Supplementary Text) reveals a distinct upturn component with a much longer time constant than the RXD reduction. This upturn signal exhibits clear resonant enhancement at the Cu $L_3$ edge (Fig.~\ref{fig:2}\textbf{E}). Pump fluence dependence at a fixed delay time after photoexcitation (4.5 ps) demonstrates that the upturn signal arises only above a certain fluence where ultrafast sublattice demagnetization occurs (Fig.~\ref{fig:2}\textbf{F}). The upturn signal decreases at high fluence because the RXD intensity reduction is larger at temperatures closer to the AFM transition temperature (Supplementary Text for temperature estimation). Crucially, the incident X-ray polarization dependence shows significantly larger upturn signals for $\uppi$ polarization than $\upsigma$ polarization (Fig.~\ref{fig:2}\textbf{G}). This polarization dependence is similar to the magnetic RXD profile (Fig.~\ref{fig:S5}) and reveals a magnetic origin of the upturn signals rather than an isotropic charge origin. Furthermore, the $(\nicefrac{1}{2}\ 0\ {-\nicefrac{1}{2}})$ RXD peak width remains nearly constant while the tail intensity rises (Fig.~\ref{fig:2}\textbf{C}), excluding the possibility of a change in the correlation length of the long-range order~\cite{Monti1} as the origin of the observed dynamics. A photo-induced phase transition into the incommensurate phase~\cite{Johnson1} is inconsistent with our observation (Supplementary Text). 

As established for resonant inelastic X-ray scattering (RIXS), the inelastic part of the resonant X-ray scattering cross-section contains information on the dynamical charge susceptibility and the dynamical spin-spin correlation function~\cite{Ament1}. The phase factor around the half-integer momentum point suppresses the dynamical charge susceptibility and low-energy phonon contributions. In contrast to time-resolved RIXS (tr-RIXS), tr-RDS measures the time (\textit{t})-dependent, energy-integrated spin-spin correlation~\cite{Haverkort1}. As detailed in Methods, the RDS intensity follows:
\begin{equation}
    I_\mathrm{RDS}(\mathbf{q}, t) \propto \sum_{\alpha \alpha^\prime} [\hat{\epsilon}_i \times \hat{\epsilon}_o]^*_\alpha [\hat{\epsilon}_i \times \hat{\epsilon}_o]_{\alpha^\prime} S_{\alpha \alpha^\prime}(\mathbf{q}, t) \ ,
    \label{eq:1}
\end{equation}
where $\hat{\epsilon}_i$ ($\hat{\epsilon}_o$) denotes the incoming (outgoing) polarization; $S_{\alpha\alpha^\prime}(\mathbf{q}, t)$ is the momentum (\textbf{q})-dependent spin-spin correlation function for each time snapshot; $\alpha$ and ${\alpha}^\prime$ denote the Cartesian coordinate directions of the lab frame. In leading order, the correlation function is determined by the magnon distribution $f_\mu(\mathbf{q},t)$~\cite{Toth1}:
\begin{equation}
    S_{\alpha \alpha^\prime}(\mathbf{q},t) \propto \sum_{\mu} \left[ A_{\mu}(\mathbf{q}) f_{\mu}(\mathbf{q}, t) + \bar{A}_{\mu}(\mathbf{q})\bar{f}_{\mu}(\mathbf{q}, t) \right]\ ,
    \label{eq:2}
\end{equation}
where $A_\mu(\mathbf{q})$ and $\bar{A}_{\mu}(\mathbf{q})$ are prefactors specific to the magnetic ground state (see Methods) and $\bar{f}_\mu = f_\mu + 1$. 
Hence, tr-RDS allows us to directly access the time- and momentum-dependent magnon distribution triggered by the ultrafast spin disorder, apart from geometric factors. 

The magnon dispersion, defined by exchange interactions, follows the magnetic structure symmetry (magnetic point group $2/m.1^{\prime}$), and therefore the RDS signals of magnon origin should also respect the symmetry. Figure~\ref{fig:3} shows the momentum dependence of the RDS intensities around the $(\nicefrac{1}{2}\ 0\ {-\nicefrac{1}{2}})$ reflection (hereafter, referred to as $\textbf{q}_{\text{RXD}}$) along $\mathbf{b}^\ast$. Although the resonant X-ray scattering signals at the momentum points close to $\textbf{q}_{\text{RXD}}$ have a sizable quasi-instantaneous intensity reduction attributed to the RXD tail, the signals also show a substantial upturn at these momentum points at later timescales (Fig.~\ref{fig:3}\textbf{A}). Fitting with two exponential functions successfully separates the RDS signals from the RXD signals (Figs.~\ref{fig:3}\textbf{C},~\ref{fig:3}\textbf{E}, and S6). While the RXD signal amplitude increases with proximity to $\textbf{q}_{\text{RXD}}$, its time constant remains constant (Fig.~\ref{fig:3}\textbf{D}), indicating a global suppression of the AFM order, i.e., ultrafast sublattice demagnetization. The amplitude of RDS signals is symmetric with respect to $\textbf{q}_{\text{RXD}}$ along $\mathbf{b}^\ast$ [Fig.~\ref{fig:3}\textbf{F} (orange)], as expected from the magnon dispersion in a centrosymmetric structure. Compared to RDS signals along $\mathbf{b}^\ast$, the amplitude of RDS signals along $[h\ 0\ h]$ significantly differs (Fig.~\ref{fig:S9}), indicating strong anisotropy in magnon dispersion as expected from the symmetry, while RXD signals are nearly identical between the two directions. 
%These observations further confirm that the upturn signals represent RDS of magnon origin. 

%Although a photo-induced phase transition into the incommensurate phase~\cite{Johnson1} could result in an upturn signal in a certain momentum range via the transient appearance of a corresponding magnetic Bragg peak, it does not explain our experimental observation in the moderate fluence regime. This is because (i) the measured momentum point $(0.49, 0.01, -0.51)$ is distant from the incommensurate peak position at $(0.506, 0, -0.483)$, (ii) it does not explain the systematic tail intensity increase around the magnetic Bragg peak (Fig.~\ref{fig:2}\textbf{C}) and (iii) the momentum-dependent time constants (Fig.~\ref{fig:3}\textbf{F}; discussed in the next section), and (iv) we did not experimentally detect any significant upturn signal at the incommensurate peak position compared to other momentum points after the photoexcitation within our experimental precision. The overall temperature increase estimated by the specific heat~\cite{Gmelin1} and deposited energy density is below 20 K ($<$ $T_\mathrm{N1}$), indicating the remaining collinear AFM state after the photoexcitation. Using a larger excitation laser fluence can lead to the transient incommensurate state with photoexcited magnons and full suppression of the commensurate order (Fig.~\ref{fig:S7}) once the average temperature in the probed volume goes above $T_\mathrm{N1}$. 

\subsection*{Quasi-thermalization after photoexcitation}

Unlike the RXD time traces, the time constant of RDS signals [Fig.~\ref{fig:3}\textbf{F} (blue)] strongly depends on momentum. This indicates a time-dependent redistribution of the magnon density in reciprocal space along the dispersion curve. Given the time scale of a few picoseconds, the likely primary mechanism behind the observed magnon redistribution is magnon-magnon scattering. 
Magnon-electron scattering can be excluded since the electrons are already thermalized within femtosecond timescales~\cite{Novelli1, Baykusheva1}; magnon-phonon scattering occurs at later timescales (discussed in the next section). Considering the significant difference in spin gap energies between the acoustic ($\sim$1.3 meV) and optical ($\sim$23.2 meV) magnon branches~\cite{Aïn1, Monod1}, the observed tr-RDS signals around the $\Gamma$ point dominantly originate from the acoustic magnons and their quasi-thermalization via magnon-magnon scattering. 
%Magnon-magnon scattering arises from the magnetic interaction, particularly the exchange couplings, which determine the magnon dispersion. Hence, the momentum-dependent scattering rate ($h/\tau_{\text{RDS}}$) of the tr-RDS signals, where \textit{h} is the Planck constant, should correspond to magnon energy and is in good agreement with the previously reported acoustic magnon dispersion (Fig.~\ref{fig:3}\textbf{I}). 
%This indicates the observed tr-RDS signals around the $\Gamma$ point originate from the acoustic magnons and their quasi-thermalization via magnon-magnon scattering, as consistent with the significant difference in spin gap energies between the acoustic ($\sim$1.3 meV) and optical ($\sim$23.2 meV) magnon branches~\cite{Aïn1, Monod1}. 
Thus, tr-RDS complements tr-RIXS, as the latter accesses only high-energy magnetic excitation due to limited intrinsic energy resolution~\cite{Mazzone1, Padma1}.

To theoretically investigate the magnon dynamics, we employ a quantum-kinetic approach based on a spin-charge-coupled model derived from the Hubbard model~\cite{Huang1} (Supplementary Text). The photoexcitation is modeled as quasi-instantaneous creation of doublons, creating non-thermal magnons. With this initial state as input, we solve the quantum Boltzmann equation for magnons,
\begin{align}
	\label{eq:QBE}
	\frac{d}{dt} f_\mu(\mathbf{q},t) = S_\mu[\{f\}](\mathbf{q},t) \ ,
\end{align}
where $S_\mu[\{f\}](\mathbf{q},t)$ is the magnon-magnon scattering integral (similar to electron-electron scattering) derived from the spin Hamiltonian in leading order beyond linear spin-wave theory. Our calculations are performed for a simplified orthorhombic unit cell (Supplementary Text) to reduce computational costs while faithfully reproducing the low-energy magnon dynamics. Combining the calculated time-dependent magnon distribution $f_\mu(\mathbf{q},t)$ with Eqs.~\ref{eq:1} and~\ref{eq:2} enables us to simulate the tr-RDS signals.

Our simulations reveal that the magnons created via the spin-charge coupling are initially distributed relatively broadly throughout the Brillouin zone, redistribute energy through the magnon-magnon scattering, and then accumulate around the $\Gamma$ point (Figs.~\ref{fig:3}\textbf{J}--3\textbf{N} and~\ref{fig:magnon_evol}), where the acoustic magnon energy is minimum with a spin gap~\cite{Aïn1, Monod1}. Note that the quasi-instantaneous spin-charge coupling is only weakly momentum dependent, and thus the initial magnon distribution is non-thermal (Fig.~\ref{fig:magnon_occupation}), leading to momentum-dependent hot magnon temperatures, similar to previous studies on a ferromagnet~\cite{Weißenhofer1}. The scattering rate ($1/\tau_{\text{RDS}}$) is proportional to momentum (Fig.~\ref{fig:3}\textbf{I}). Hydrodynamic magnon slowdown via magnon-magnon scattering was reported in antiferromagnets~\cite{Harris1}, and our simulations also show the slowdown (Fig.~\ref{fig:time_evol_cuts}). 
The calculated tr-RDS profile based on the time evolution of magnon density distribution (Fig.~\ref{fig:3}\textbf{H}) reproduces the momentum-delay time plots of the upturn component of the resonant X-ray scattering intensities (compare Figs.~\ref{fig:3}\textbf{E} and~\ref{fig:3}\textbf{G}). The magnon-magnon scattering changes the magnon distribution while keeping the total magnon energy constant and reaches a quasi-thermalized state within the branch fulfilling Bose-Einstein statistics. While similar magnon redistribution behavior was previously implied in Mott insulators using tr-RIXS without detailed time-momentum mapping~\cite{Mazzone1}, our tr-RDS measurements together with the fully quantum-mechanical calculations provide the first direct visualization of this microscopic quasi-thermalization process in reciprocal space. Energy integration in tr-RDS enables momentum-space mapping with higher photon counts than energy-resolved tr-RIXS, facilitating direct visualization of real-time momentum-resolved quasi-particle dynamics.

%Note that the total magnon density may not be conserved at magnon-magnon scattering in an AFM material because of the mixed ground state of different total spin values caused by the SU(2) symmetry breaking\cite{Zhitomirsky1}. The magnon-magnon scattering process across non-thermal to quasi-thermal distributions within the acoustic magnon branch explains the complex, non-monotonic RDS time traces observed at $(0.48, 0.01, -0.52)$ shown in Fig.~\ref{fig:S8}. 

%The non-thermally distributed magnon densities allow us to estimate corresponding "temperatures" as a function of momentum. When an excitation predominantly couples to a specific mode, the phenomenological temperature model requires subdividing the subsystem: for example, hot optical phonons in the lattice\cite{Perfetti1}. In our case, the spin system is subdivided into hot acoustic magnons. Hot-magnon temperatures at fast timescales reach $\sim$500 K at $(0.5, 0.1, -0.5)$, far exceeding the bulk quasi-equilibrium limit of $\sim$210 K. This indicates that the excitation energy initially deposited in the electronic system, resulting in doublon creation that quickly decays, transfers into the magnonic subsystem at ultrafast timescales, while the electronic states return to equilibrium~\cite{Cilento1, Baykusheva1} and the average temperature, except for hot carriers, remains below $T_\mathrm{N1}$. 

\subsection*{Magnetic order recovery via magnon-phonon scattering}

After quasi-thermalization through magnon-magnon scattering, the suppressed AFM order starts to recover at significantly later timescales with a $\sim$7 ns time constant, as diagnosed from the restoration of the RXD intensity (Fig.~\ref{fig:4}\textbf{D}, green). The RDS intensity at $(0.47, 0, -0.53)$ shows a similar time constant (orange) to the RXD time trace. This demonstrates that magnon annihilation restores the AFM ground state. 
%We hereafter discuss a local recovery process. 
%further validating the magnon origin of RDS signals. Besides, this direct correlation between RXD and RDS signals indicates that  

The spin gap at the $\Gamma$ point prevents magnons from full relaxation solely via magnon-magnon scattering. To fully recover the AFM order, phonons provide the most plausible energy sink for magnon annihilation. Supporting this relaxation mechanism, tr-NRDS intensity around the $(0\ 0\ {-2})$ Bragg reflection collected in the AFM phase exhibits a broad enhancement around 2 ns, a similar timescale to the RDS signals (Fig.~\ref{fig:4}\textbf{D}, blue), indicating phonon creation during the magnon annihilation process. The occurrence of this process suggests that the magnon diffusion process is inefficient, likely due to the magnon accumulation around the $\Gamma$ point after quasi-thermalization. This enhancement is absent in the paramagnetic phase (red), where magnon annihilation and phonon creation are unnecessary for relaxation since spins are already thermally randomized. Note that this enhancement is not caused by a shift of the Bragg peak due to thermal expansion (Fig.~\ref{fig:S10}\textbf{C}). The short-lived peak with a $\sim$160 ps time constant observed in both phases immediately after excitation arises from photo-created phonons~\cite{Trigo2} (Fig.~\ref{fig:1}\textbf{A}) or via magnon-number conserved scattering (see below) and provides the phonon quasi-thermalization timescale. The tr-NRDS signals arise predominantly where equilibrium phonons exist  (Figs.~\ref{fig:S10}\textbf{A} and~\ref{fig:S10}\textbf{C}), consistent with the faster phonon quasi-thermalization timescale compared to the phonon creation timescale via magnon annihilation.

There are three major pathways to decay energies from magnons into phonons based on a harmonic oscillator coupled to magnons\cite{Streib1} (Supplementary Text): (i) the magnon-phonon interconversion process at their crossing point (Fig.~\ref{fig:4}\textbf{A}), (ii) number-conserving magnon-phonon scattering, where a magnon loses energy by creating a phonon (Fig.~\ref{fig:4}\textbf{B}), and (iii) the anomalous scattering process, where two magnons combine into a single phonon (Fig.~\ref{fig:4}\textbf{C}). The RDS decay (Fig.~\ref{fig:4}\textbf{D}, orange) demonstrates that magnon annihilation occurs on nanosecond timescales. Of the three lowest-order magnon-phonon coupling processes, only (i) and (iii) can annihilate magnons, while (ii) conserves magnon number, likely contributing to the short-lived peak in tr-NRDS signals for quasi-thermalization. Efficient scattering requires phonon energies comparable to magnon energies, severely restricting the available phase space due to energy and momentum conservation. (i) is especially inefficient because the reciprocal space volume occupied by the intersecting points is tiny. Besides the spin gap and irrelevant magnon diffusion, the inefficient coupling strengths impose an intrinsic bottleneck for the magnetic recovery. 
%Note that although the process (ii) could contribute to the magnon quasi-thermalization process, the faster phonon relaxation after time zero than magnon quasi-equilibration (compare upper and lower panels of Fig.~\ref{fig:4}\textbf{I}) indicates its smaller impact on the quasi-thermalization process than magnon-magnon scattering. 

To elucidate the dominant magnon-phonon coupling mechanism, we calculated the respective scattering rates based on the quantum Boltzmann equation (Supplementary Text), assuming the quasi-thermalized magnon distribution after magnon-magnon scattering as the initial state. The anomalous process between the lowest-energy optical phonons and the acoustic magnons is found as the dominant magnon decay channel [see Fig.~\ref{fig:4}\textbf{E} (phonons) and 4\textbf{F} (magnons) for their scattering rates; see also Fig.~\ref{fig:magnon_phonon_scattering} and Supplementary Text]. 
The characteristic pattern is due to the phase-space constraints for the scattering process. The effective magnon-phonon coupling strength via the anomalous scattering estimated from the magnon relaxation timescale (Fig.~\ref{fig:4}\textbf{D}) is $\sim$59 $\upmu$eV, while spin-phonon coupling in CuO can be as large as $\sim$6.2 meV\cite{Chen1}. Even though spin-lattice coupling strength in CuO is expected to be generally large, as manifested by multiferroicity~\cite{Kimura1} and their persistence correlation at high temperatures\cite{Zheng1}, the strong restriction of the available phase space results in the unexpectedly small effective magnon-phonon coupling strength. This suggests the possibility of designing the lifetime of photo-induced intermediate states by exploiting corresponding quasiparticle dispersions. 

%However, in our case, all the processes are inefficient with their scattering rates of \hl{XX$^{-1}$ ns, XX$^{-1}$ ns, and XX$^{-1}$ ns}, respectively. The respective reasons are; (i) crossing points between magnon and phonon dispersions exist only at high energies ($>$ 16.5 meV; Fig. \hl{SX})~\cite{Reichardt1, Aïn1} (ii) only acoustic phonons can have lower energies than magnons but their dispersion slopes are significantly different, making energy and momentum conservations during this process nearly impossible; and (iii) some optical phonons can satisfy the momentum and energy conservation laws at the process but only at high energies ($>$ 16.5 meV), resulting in indirect scattering for low-energy magnons. Additionally, the strongly dispersive magnon branch makes anomalous scattering relevant only in the vicinity of the $\Gamma$ point, where the coupling strength intrinsically becomes insignificant. 

%(i) the direct conversion process is strongly momentum-selective as it requires magnon-phonon band crossing; 

\subsection*{Conclusions}

\begin{figure}
	\centering
	\includegraphics[width=0.9\textwidth]{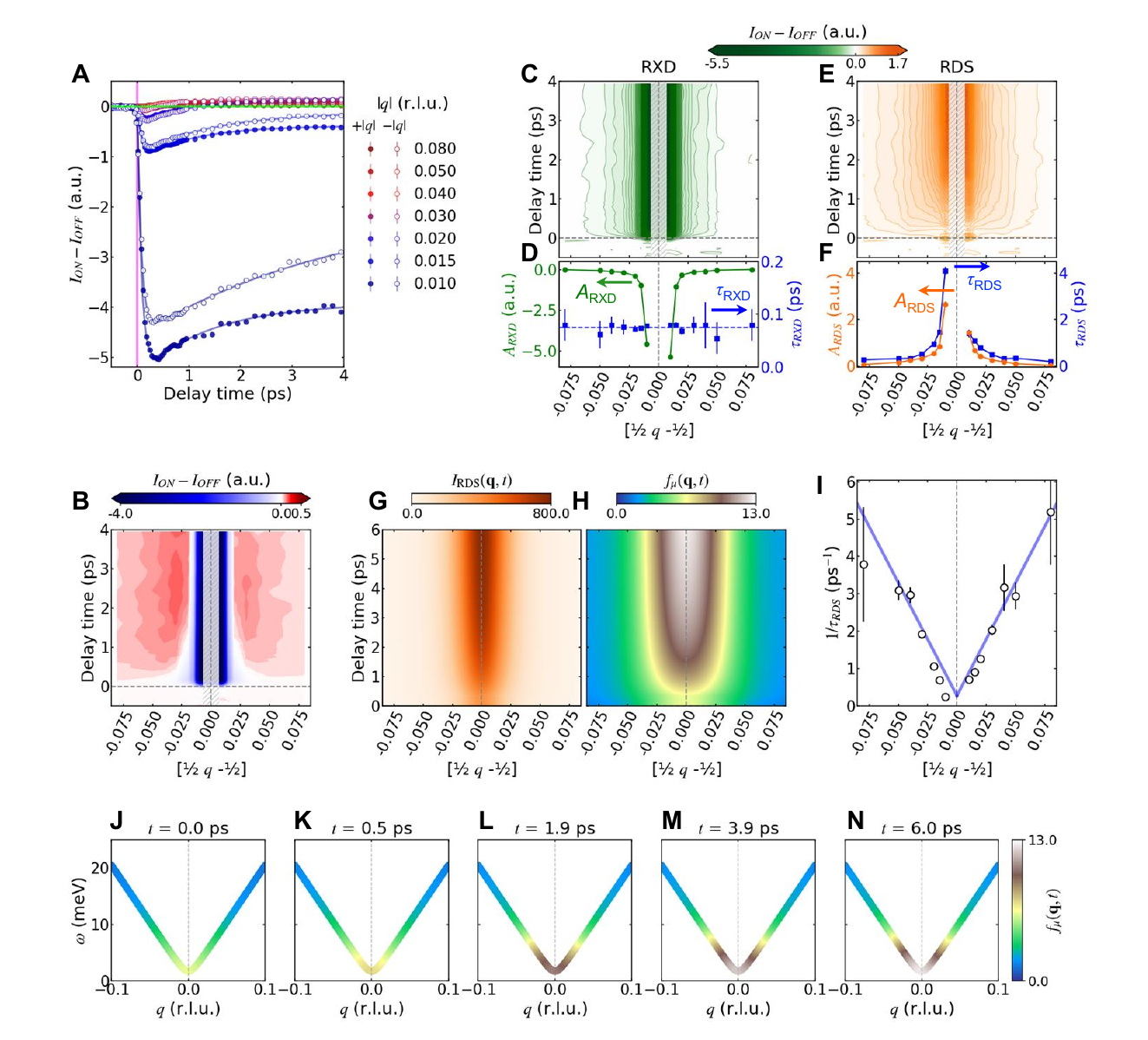} 
	% Pick an appropriate width - in print, figures are usually one or two columns wide, which can
	% be approximated by 0.3\textwidth or 0.6\textwidth respectively. Use appropriate label sizes.
	\caption{\textbf{Momentum dependence of time-resolved resonant diffuse scattering.}}
    \label{fig:3}
\end{figure}

\begin{figure}
	\centering
	\includegraphics[width=0.9\textwidth]{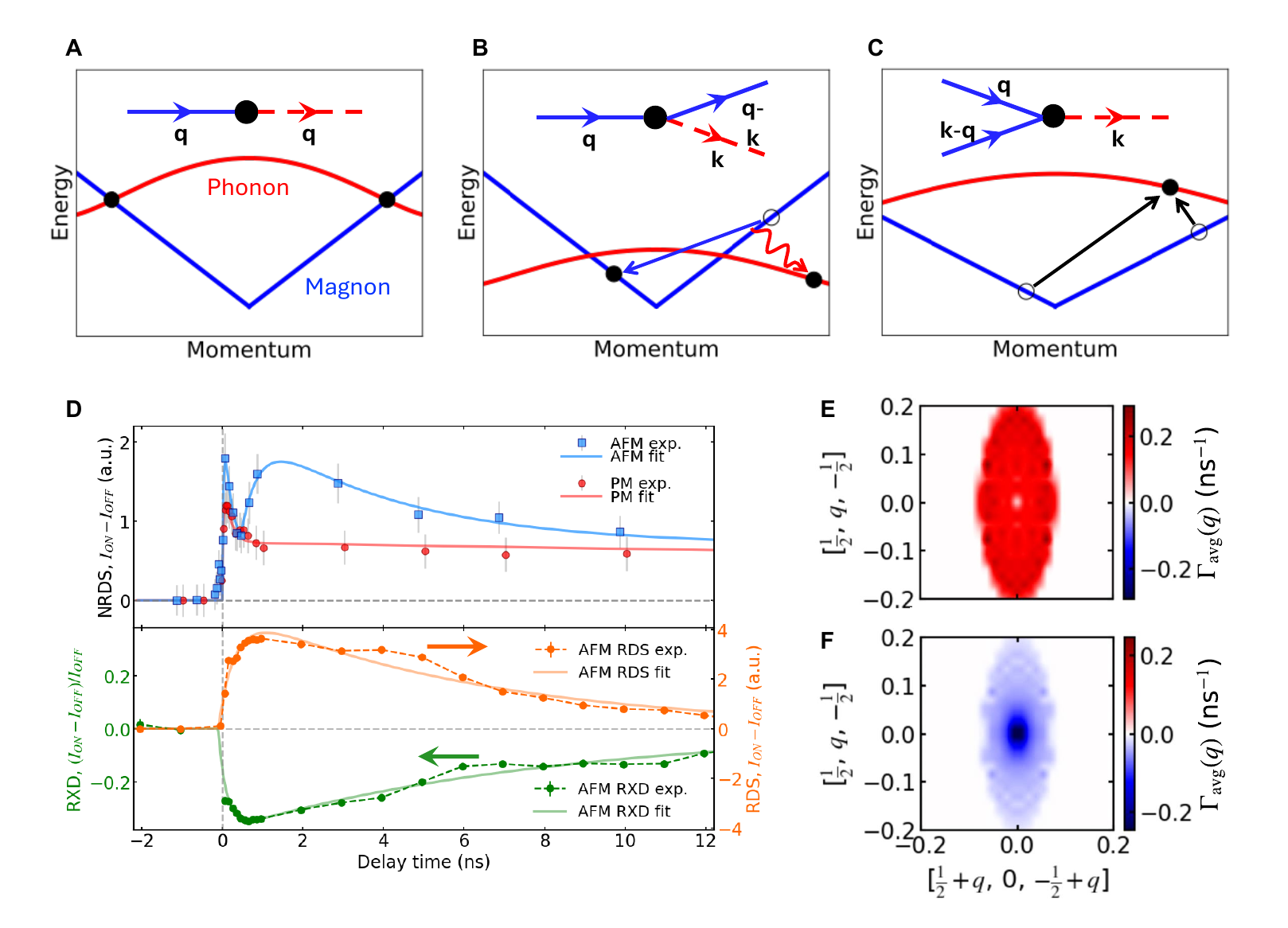} 
	% Pick an appropriate width - in print, figures are usually one or two columns wide, which can
	% be approximated by 0.3\textwidth or 0.6\textwidth respectively. Use appropriate label sizes.
	\caption{\textbf{Energy transfer from magnons to phonons.}
		(\textbf{A-C}) Possible magnon-phonon coupling mechanisms: (\textbf{A}) magnon-phonon interconversion, (\textbf{B}) magnon number-conserved scattering, and (\textbf{C}) anomalous scattering, where two magnons convert into a phonon. Insets show corresponding Feynman diagrams, where \textbf{q} and \textbf{k} represent the momentum of quasiparticles. (\textbf{D}) Long timescale comparison of the $(\nicefrac{1}{2}\ 0\ {-\nicefrac{1}{2}})$ RXD intensity (green), RDS intensity at $(0.47, 0, -0.53)$ (orange) (bottom), and NRDS intensities integrated around the $(0\ 0\ {-2})$ Bragg peak within 0.1 r.l.u. (Supplementary Text) in the collinear AFM (blue) and PM (red) phases (top), revealing a correlation between magnon annihilation and phonon creation during magnetic recovery. The used pump laser fluence was 3 mJ/cm$^{2}$. Error bars represent standard error. (\textbf{E, F}) Simulated two-dimensional distribution patterns of the scattering rate for the lowest-energy optical phonons (\textbf{E}) and the acoustic magnons (\textbf{F}) via the anomalous scattering. Here, we use the simplified orthorhombic unit cell, as mentioned in the text.}
%        Fluence-dependent spin temperatures determined by the lost sublattice magnetization and calculated magnon dispersion. Dashed curves show temperatures assuming complete energy transfer to the entire system using the total specific heat (black) or the spin system using the magnetic specific heat (gray). The transition temperatures are indicated as an eye guide. 
        
	\label{fig:4}
\end{figure}

In summary, our observations establish the complete hierarchical energy transfer pathways in non-equilibrium states associated with ultrafast sublattice demagnetization in AFM CuO. Although respective timescales may differ depending on scattering efficiencies, the processes are generalizable to any gapped magnetic insulators, including ferromagnets when they break the SU(2) symmetry, beyond multi-temperature models~\cite{Beaurepaire1}. (1) At first, photoexcitation creates spin disorder via a high-energy electronic configuration, which quickly recovers on femtosecond timescales, leaving significant excitation energy in the spin sector. This results in non-thermal low-energy magnons distributed throughout the Brillouin zone (Fig.~\ref{fig:3}\textbf{J}), alongside photo-created phonons. 
%, which may be axial phonons in ferromagnets due to angular momentum conservation~\cite{Tauchert1}. 
(2) Magnon-magnon scattering drives the magnetic system into quasi-equilibrium on picosecond timescales, with magnons accumulating towards the spin gap (Figs.~\ref{fig:3}\textbf{J}--3\textbf{N}), alongside possible magnon decays from high-energy branches and magnon-conserving scattering with phonons. (3) Magnon-phonon coupling releases energy from the spin system to phonons and recovers the magnetic order on nanosecond timescales via the anomalous scattering, annihilating magnons (Fig.~\ref{fig:4}\textbf{F}). The principle that dispersion overlap between magnons and phonons controls coupling efficiency applies generally to control the lifetime of non-equilibrium states by engineering the dispersions. (4) Phonon annihilation recovers the equilibrium state by transporting energy away on microsecond timescales (Fig.~\ref{fig:S11}). This process may be dominated by electron-phonon scattering rather than phonon diffusion in CuO because of the temperature dependence of the phonon annihilation time constant (Supplementary Text). 

%Although multi-temperature models phenomenologically describe thermalization through various couplings~\cite{Beaurepaire1}, 
Direct observation of these processes remained experimentally challenging, as it requires sensitivity to multiple subsystems with momentum resolution at ultrafast timescales. The time- and momentum-resolved view of magnon and phonon dynamics addresses the fundamental questions of ultrafast magnetism, crucial for realizing high-speed, high-repetition devices: (1) energy transfer pathways, (2) subsystem coupling efficiencies, and (3) dynamical bottlenecks. The established general framework predicts that materials with strongly overlapping dispersions of magnons and phonons will exhibit faster recovery, enabling rational design of device operation speeds. 

While tr-NRDS accesses low-energy lattice excitations~\cite{Trigo1}, tr-RDS emerges as a transformative tool to investigate low-energy electronic counterparts in non-equilibrium states. This new capability complements tr-RIXS, which accesses a variety of high-energy fundamental excitations, including magnons, orbital excitations, excitons, plasmons, and polarons~\cite{Mitrano2}, indicating the applicability of tr-RDS to a wide class of materials~\cite{Osborn1}, including superconductors~\cite{Mitrano1}, quantum paraelectrics~\cite{Li2}, and systems controlled via Floquet engineering~\cite{Ikeda1}. Low-energy quasiparticles often dominate physics on various timescales, including phase transitions, transport phenomena, and non-equilibrium states. The ability to track low-energy quasiparticle kinetics across multiple subsystems in real time provides unprecedented insight into non-equilibrium states and energy flow pathways, and determines mode-selective coupling strength. 
%design principles to stabilize long-lived non-equilibrium states by enabling the determination of mode-selective coupling strength. 
Its independence from Fourier-transform limits due to energy integration enables, in principle, attosecond-scale investigations of electronic quantum dynamics. Our framework establishes the foundation for a new generation of ultrafast control protocols, bridging fundamental condensed matter physics with transformative quantum technologies. 

\pagebreak

\begin{figure}
	\centering
	\includegraphics[width=0.9\textwidth]{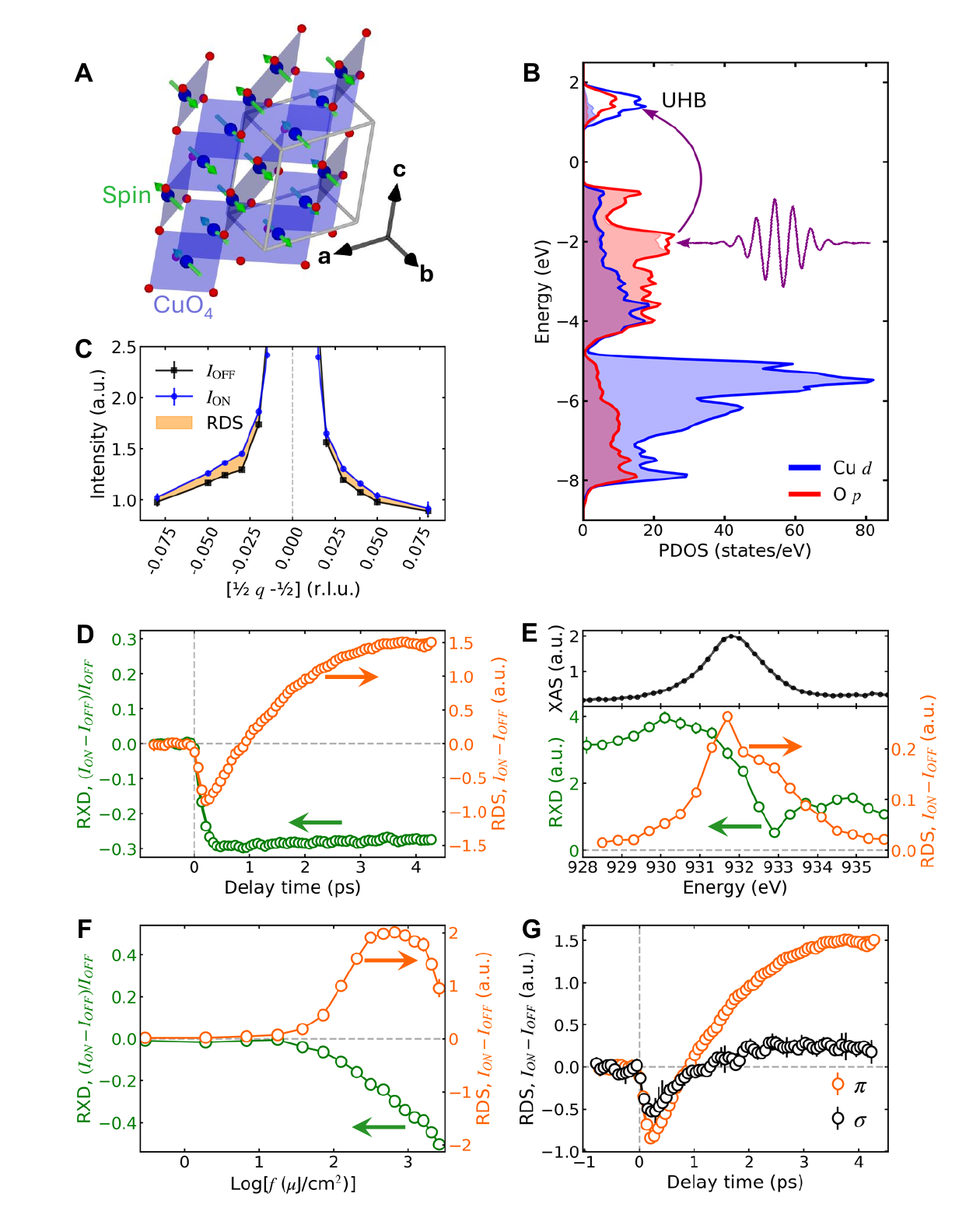} 
	% Pick an appropriate width - in print, figures are usually one or two columns wide, which can
	% be approximated by 0.3\textwidth or 0.6\textwidth respectively. Use appropriate label sizes.
    \caption{\textbf{Resonant diffuse scattering from non-equilibrium magnons.}}
	\label{fig:2}
\end{figure}
\clearpage
\noindent
(\textbf{A}) Magnetic structure of CuO in the collinear AFM phase. (\textbf{B}) Calculated projected density of states (PDOS) of CuO, showing the charge-transfer excitation by 400 nm laser pulses from O 2\textit{p} band to Cu 3\textit{d} upper Hubbard band (UHB). (\textbf{C}) Comparison of resonant X-ray scattering profiles around the $(\nicefrac{1}{2}\ 0\ {-\nicefrac{1}{2}})$ RXD peak before (black) and 3 ps after photoexcitation (blue). The profiles were extracted from the tr-RDS data shown in Fig.~\ref{fig:3}\textbf{A}. The top of the diffraction peak is missing due to saturation of the detector. (\textbf{D}) Simultaneously collected time traces of the $(\nicefrac{1}{2}\ 0\ {-\nicefrac{1}{2}})$ RXD intensity (green) and RDS intensity at $(0.49, 0.01, -0.51)$ (orange). (\textbf{E}) X-ray absorption spectrum (XAS, black), and fixed-momentum incident X-ray energy dependences of the $(\nicefrac{1}{2}\ 0\ {-\nicefrac{1}{2}})$ RXD intensity (green) and RDS intensity at $(\nicefrac{1}{2}, -0.08, {-\nicefrac{1}{2}})$ without intensity reduction due to the diffraction tail (orange). The large dip in RXD reflects the self-absorption effect~\cite{Scagnoli1}. (\textbf{F}) Laser fluence ($f$) dependence of the RXD (green) and RDS [orange, $(0.49, 0.01, -0.51)$] intensities measured at 4.5 ps after the photoexcitation. (\textbf{G}) Incident X-ray polarization dependence of RDS time traces for $\uppi$ (orange) and $\upsigma$ (black) polarizations. Error bars represent standard error. All the measurements were performed at 190 K. The pump laser fluence for data shown in \textbf{C}, \textbf{D}, \textbf{E}, and \textbf{G} was 3 mJ/cm$^{2}$. 
    
\clearpage
\noindent
(\textbf{A}) Time traces of resonant X-ray scattering intensities at momentum points around $\textbf{q}_{\text{RXD}}$ (see text) along $\mathbf{b}^\ast$, collected with the pump laser fluence of 3 mJ/cm$^{2}$. Solid curves represent fits using two exponential functions to separate the RXD reduction and RDS upturn components (Supplementary Text). Filled (open) circles indicate data measured at positive (negative) momentum deviation $q$ from $\textbf{q}_{\text{RXD}}$. (\textbf{B}) Momentum-delay time contour plots of the experimental data shown in \textbf{A}. (\textbf{C, E}) Momentum-delay time contour plots of the RXD (\textbf{C}) and RDS (\textbf{E}) components derived from the time traces in \textbf{A} by decomposing into the two components. (\textbf{D, F}) Momentum dependence of RXD-reduction (\textbf{D}) and RDS-upturn (\textbf{F}) amplitudes (\textit{A}) and their time constants ($\tau$) extracted from fits to the data in \textbf{A}, demonstrating the symmetric amplitude profile expected from the symmetry and magnon redistribution. Error bars represent standard error in \textbf{A} and are from the fit in \textbf{D} and \textbf{F}. The hatched area represents the momentum range where strong RXD signals prevented us from measuring resonant X-ray scattering intensities due to saturation. (\textbf{G}) Simulated RDS time traces and (\textbf{H}) time-dependent magnon distribution along $[\nicefrac{1}{2}\ q\ {-\nicefrac{1}{2}}]$. (\textbf{I}) The scattering rate $1/\tau_{\text{RDS}}$ as a function of momentum along $[\nicefrac{1}{2}\ q\ {-\nicefrac{1}{2}}]$. Solid lines are guide to the eyes. (\textbf{J--N}) Magnon density distribution along $[\nicefrac{1}{2}\ q\ {-\nicefrac{1}{2}}]$ at (\textbf{J}) $t$ = 0 ps, (\textbf{K}) 0.5 ps, (\textbf{L}) 1.9 ps, (\textbf{M}) 3.9 ps, and (\textbf{N}) 6.0 ps, derived from tr-RDS data using Eq. \ref{eq:2}.

%%%%%%%%%%%%%%%% MAIN TEXT TABLES %%%%%%%%%%%%%%%

% \begin{table} % Do NOT use \begin{table*}
% 	\centering
% 	% Captions go above tables
% 	\caption{\textbf{All captions must start with a short bold sentence, acting as a title.}
% 		Then explain what is being listed in the table, the meaning of each column etc.
% 		Captions are placed above tables.}
% 	\label{tab:example} % give each table a logical label name
	
% 	\begin{tabular}{lccc} % four columns, alignment for each
% 		\\
% 		\hline
% 		Sample & $A$ & $B$ & $C$\\
% 		 & (unit) & (unit) & (unit)\\
% 		\hline
% 		First & 1 & 2 & 3\\
% 		Second & 4 & 6 & 8\\
% 		Third & 5 & 7 & 9\\
% 		\hline
% 	\end{tabular}
% \end{table}

%%%%%%%%%%%%%%%% REFERENCES %%%%%%%%%%%%%%%

\clearpage % Clear all remaining figures and tables then start a new page

% The list of references goes after the main text and before the acknowledgements
% When preparing an initial submission, we recommend you use BibTeX, like this:
%
\bibliography{science_template} % for a file named science_template.bib
\bibliographystyle{sciencemag}

% After the paper has completed peer review and been revised ready for acceptance,
% you should comment out the lines above and copy-paste the contents of your .bbl
% file here instead. This will help ensure that our conversion software works correctly.
% Remember to re-run BibTeX first - check the timestamp!
%
% Example of the first three entries copy-pasted from science_template.bbl:
%
%\begin{thebibliography}{1}
%
%\bibitem{example}
%A.~N. {Author}, An example reference. \emph{Journal of Improbable Research}
%  \textbf{1}, 67 (2020).
%
%\bibitem{example2}
%F.~M. {Surname}, S.~{Author}, A second example. \emph{Interesting Research
%  Letters} \textbf{32}, 897 (2019).
%
%\bibitem{example_preprint}
%P.~{One}, P.~{Two}, P.~{Three}, {An unpublished preprint}. \emph{preprint}
%  (2021), arXiv:2101.12345.
%
%\end{thebibliography}

%%%%%%%%%%%%%%%% ACKNOWLEDGEMENTS %%%%%%%%%%%%%%%

\section*{Acknowledgments}
The time-resolved resonant X-ray diffraction and resonant diffuse scattering experiments were performed at the Furka endstation of the Athos branch of SwissFEL at Paul Scherrer Institute as in-house beamtime. The time-resolved non-resonant diffuse scattering experiment was conducted at the ID09 beamline of ESRF under proposal No. HC-6207. 
This work was supported by a grant from the Swiss National Supercomputing Centre (CSCS) under projects "psi10", "mr34" on Alps.

\paragraph*{Funding:}
PSI Research Grant 2022 (ARC, MCH, ER, HU);
NCCR MARVEL (KSM, MS); 
SNSF Ambizione Grant PZ00P2-193527 (YY, MS);
Marie Sklodowska-Curie Grant 884104 (AN)

%
%This research was supported by the NCCR MARVEL, a National Centre of Competence in Research, funded by the Swiss National Science Foundation (grant number 205602). Y.Y and M.S. acknowledge support from SNSF Ambizione Grant No. PZ00P2-193527.

\paragraph*{Author contributions:} Conceptualization: HU \\
Sample preparation: ARC, EPo, and MCH \\
Methodology: US, MS, ER, and HU \\
Investigation: ARC, ES, YY, BL, AN, SWH, LL, KSM, GP, SZ, RM, CM, ML, EPa, LP, US, MS, ER, and HU \\
Visualization: ARC, MS, and HU \\
Funding acquisition: AN, MCH, MS, ER, and HU \\
Writing – original draft: ARC, MS, and HU \\
Writing – review \& editing: ARC, ES, AN, SZ, EPa, LP, US, MS, ER, and HU
\paragraph*{Competing interests:}
The authors declare no competing interests.
\paragraph*{Data and materials availability:}
All data needed to evaluate the conclusions are available in the main text or the supplementary materials and accessible at the PSI Data Repository~\cite{PSIDataRepository}. 
\subsection*{Supplementary materials}
Materials and Methods\\
Supplementary Text\\
Figs.~\ref{fig:S1} to~\ref{fig:magnon_phonon_scattering}\\
References \textit{(51-\arabic{enumiv})}\\ % automatically fills out the last reference number
% (filling out the other numbers automatically is possible but fiddly and liable to break)

%%%%%%%%%%%%%%%% END OF MAIN TEXT %%%%%%%%%%%%%%%

\newpage

%%%%%%%%%%%%%%%% START OF SUPPLEMENT %%%%%%%%%%%%%%%

% Figures, tables, equations and pages in the supplement are numbered S1, S2 etc.
\renewcommand{\thefigure}{S\arabic{figure}}
\renewcommand{\thetable}{S\arabic{table}}
\renewcommand{\theequation}{S\arabic{equation}}
\renewcommand{\thepage}{S\arabic{page}}
\setcounter{figure}{0}
\setcounter{table}{0}
\setcounter{equation}{0}
\setcounter{page}{1} % not 0 as \newpage already started a supplementary page
% References continue the numbering from the main text.

%%%%%%%%%%%%%%%% SUPPLEMENT TITLE PAGE %%%%%%%%%%%%%%%

\begin{center}
\section*{Supplementary Materials for\\ \scititle}

% Author list for the supplement
% Indicate the corresponding authors, but do NOT include institutions here
% It would be nice if the template auto-generated this, but doing so is complicated...
Arnau~Romaguera, Elizabeth~Skoropata, Yun~Yen, Biaolong~Liu, Abhishek~Nag, Shih-Wen~Huang, Ludmila~Leroy, 
Katja~Sophia~Moos, Gian~Parusa, Serhane~Zerdane, Ritwika~Mandal, Céline~Mariette, Matteo~Levantino, Eugenio~Paris,  Luc~Patthey, 
Ekaterina~Pomjakushina, Urs~Staub, Monica~Ciomaga~Hatnean, Michael~Schüler, Elia~Razzoli, and Hiroki~Ueda$^{\ast}$\\
\small$^\ast$Corresponding author. Email: hiroki.ueda@psi.ch\\
% \small$^\dagger$These authors contributed equally to this work.
\end{center}

% Fill out the numbers for each type of supplementary material,
% and delete any lines that aren't applicable.
% These are just example numbers that don't match the rest of this template.
\subsubsection*{This PDF file includes:}
Materials and Methods\\
Supplementary Text\\
Figs.~\ref{fig:S1} to \ref{fig:magnon_phonon_scattering}\\
References\\

\newpage

%%%%%%%%%%%%%%%% MATERIALS AND METHODS %%%%%%%%%%%%%%%

\subsection*{Materials and Methods}

\subsubsection*{\underline{Experimental setup of tr-RXS and tr-RDS}}

\begin{figure} 
	\centering
	\includegraphics[width=0.7\textwidth]{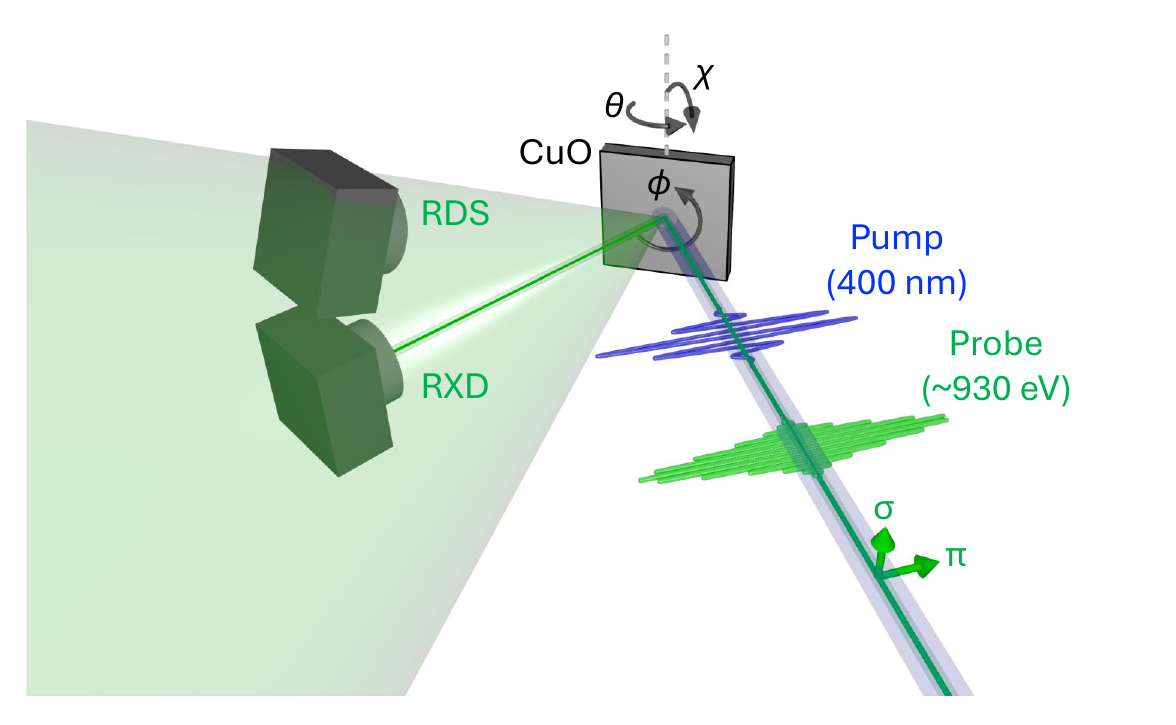}
	\caption{\textbf{Schematic of experimental setup for tr-RXD and tr-RDS measurements.}
		}
	\label{fig:S1}
\end{figure}

The experimental geometry for time-resolved resonant X-ray diffraction (tr-RXD) and time-resonant diffuse scattering (tr-RDS) measurements, performed at the Furka endstation of the SwissFEL at Paul Scherrer Institute, is illustrated in Fig.~\ref{fig:S1}. A single-crystal CuO sample, oriented with its largest surface parallel to the $(1\ 0\ {-1})$ plane, was mounted on a four-circle diffractometer equipped with three sample rotation stages ($\theta$, $\varphi$, $\chi$) and two detector rotation stages (2$\theta$) in an ultra-high vacuum chamber. Signal detection was achieved using a photodiode for RXD measurements and two reverse-biased avalanche photodiodes for RDS measurements. The active areas of all the detectors were masked with a thin aluminium foil to reject optical photons. The photodiode and one of the avalanche photodiodes were mounted on a single tower assembly to enable simultaneous data acquisition at nearby momentum points. While the photodiode was placed on the horizontal plane, the avalanche photodiode mounted on the same tower was slightly above it. The other avalanche photodiode was on the horizontal plane. For the investigation of RXD at high excitation fluence, we used a 0.5M Jungfrau detector. Note that tr-RDS signals were detected only with avalanche photodiodes due to their high quantum efficiency. X-ray absorption spectrum was measured with an avalanche photodiode in the total fluorescence yield mode. The sample temperature was fixed at 190 K using a cryostat. 

The optical pump pulses were generated via second-harmonic generation of fundamental Ti: sapphire laser pulses (800 nm) through a 150 $\upmu$m-thick $\upbeta$-BaB$_2$O$_4$ crystal, providing 400 nm pulses with \textit{p}-polarization, $\sim$35 fs pulse duration, and 100 Hz repetition rate. The pump beam was nearly collinear relative to the probe beam using an in-vacuum mirror positioned adjacent to the probe beam path. “Off” shots were collected for normalization by introducing a 28 ns delay to one out of every three pump pulses. The penetration depth of the pump beam is estimated as $\sim$450 nm~\cite{Sukhorukov1}. 

X-ray probe pulses were delivered along the Athos beamline of the SwissFEL~\cite{Prat1}, utilizing self-amplified spontaneous emission with photon energies tuned to the Cu $L_3$ edge ($\sim$930 eV) monochromatized with a bandwidth of $\sim$0.5 eV. The pulse duration and repetition rate were $\sim$50 fs and 100 Hz, respectively. X-ray polarization was controlled between $\upsigma$ (vertical) and $\uppi$ (horizontal) using the Apple X undulators~\cite{Calvi1}. Transmission gratings were used to perform shot-by-shot photon diagnostics, i.e., measuring arrival time relative to the pump beam and incoming intensity for normalization, without interfering with tr-RXD and tr-RDS measurements. The X-ray probe depth is estimated as $\sim$18 nm on the resonance at the incidence angle ($\sim$$63.5^\circ$) satisfying the magnetic Bragg condition~\cite{Staub1}. The combined time resolution is $\sim$60 fs full-width at half-maximum. 

\subsubsection*{\underline{Experimental setup of tr-NRDS}}

\begin{figure} 
	\centering
	\includegraphics[width=0.7\textwidth]{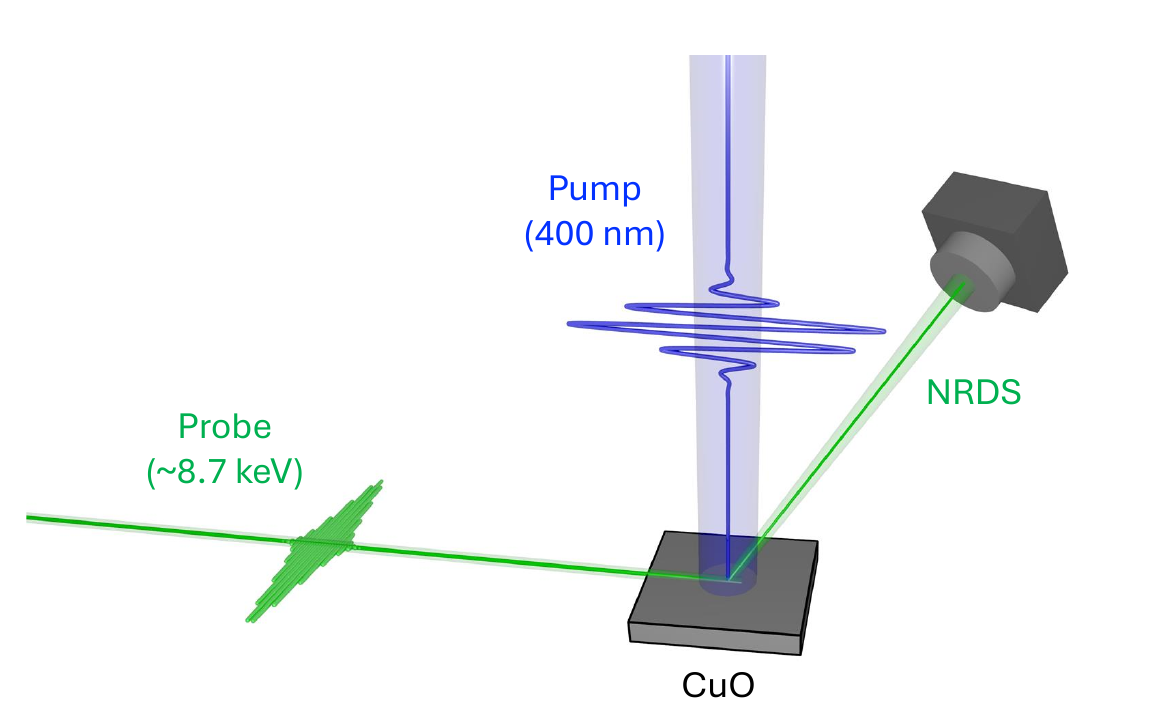}
	\caption{\textbf{Schematic of experimental setup for tr-NRDS measurements.}
		}
	\label{fig:S2}
\end{figure}

The experimental configuration of time-resolved non-resonant diffuse scattering (tr-NRDS) measurements, conducted at the ID09 beamline of the European Synchrotron Radiation Facility (ESRF), is schematically depicted in Fig.~\ref{fig:S2}. The identical single-crystal sample was mounted on a motorized goniometer with precise angular control. A 1M Jungfrau two-dimensional detector was used to collect diffuse scattering patterns. Sample temperature was controlled between 190 K and 300 K using a nitrogen cryostream. Optical pump pulses were provided by a Ti: sapphire picosecond laser (800 nm wavelength and 1 kHz repetition rate) with optical second-harmonic generation to convert the wavelength to 400 nm using a $\upbeta$-BaB$_2$O$_4$ crystal. The incidence angle of the pump beam was approximately 88$^\circ$, yielding a penetration depth of $\sim$500 nm. The X-ray probe beam was polychromatic radiation at 8.7 keV below the Cu $K$ edge to suppress fluorescence signals. Grazing incidence geometry ($\sim$2$^\circ$) was employed for the probe beam, resulting in the probe depth of $\sim$1.5 $\upmu$m in the vicinity of the $(0\ 0\ {-2})$ Bragg condition. X-ray pulses with $\sim$100 ps pulse duration and $\uppi$ polarization were synchronized to the pump laser through a mechanical chopper system~\cite{Levantino1}, reducing the repetition rate to 1 kHz to match the pump laser. 

\subsubsection*{\underline{Sample preparation and characterization}}

\begin{figure} 
	\centering
	\includegraphics[width=0.6\textwidth]{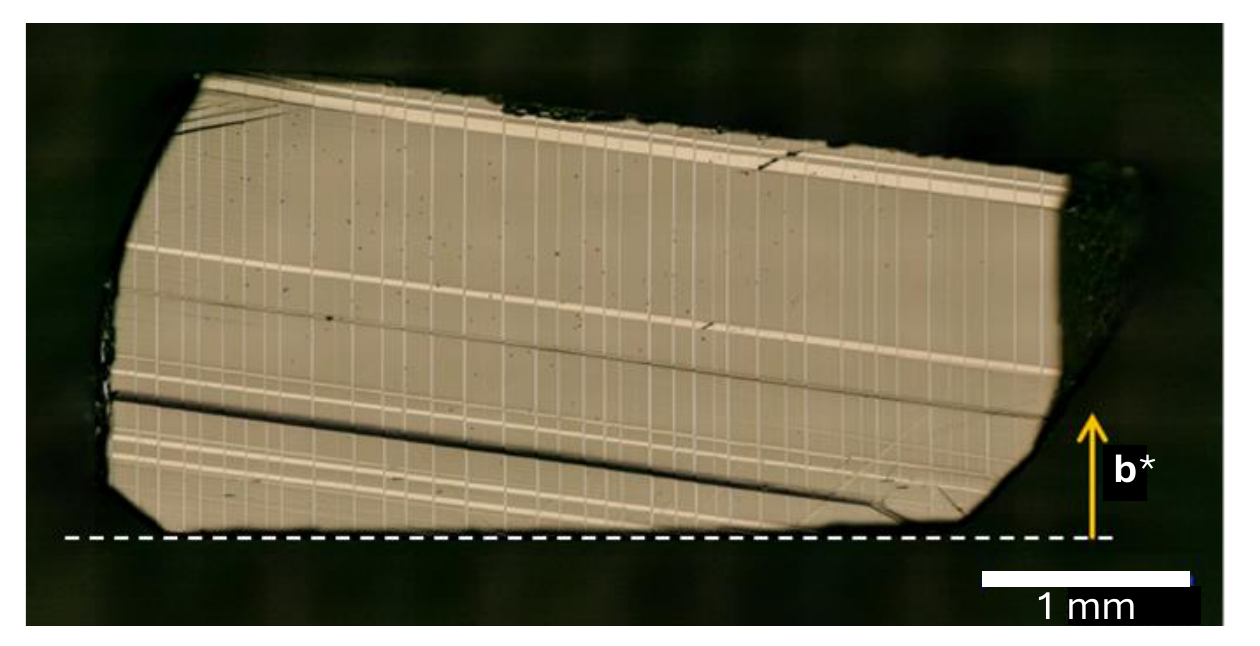}
	\caption{\textbf{Polarized microscope image on the (1 0 $\mathbf{-}$1) surface of the CuO single crystal sample.}
        Bright regions represent differently oriented twin domains.
		}
	\label{fig:S4}
\end{figure}

\begin{figure} 
	\centering
	\includegraphics[width=0.8\textwidth]{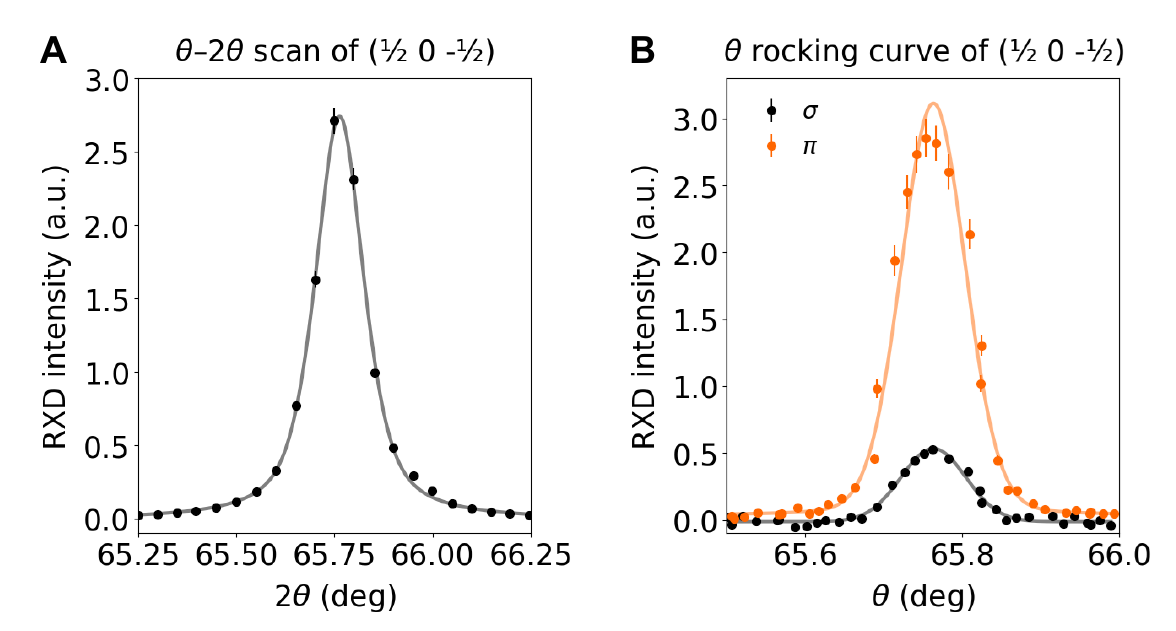}
	\caption{\textbf{Diffraction profile and rocking curve around the ($\mathbf{\nicefrac{1}{2}}$ 0 $\mathbf{-\nicefrac{1}{2}}$) magnetic Bragg reflection.}
        Polarization dependence is also shown in \textbf{B}. Solid curves represent Gaussian fit. Error bars are standard errors.
		}
	\label{fig:S5}
\end{figure}

A single crystal of CuO was grown using the floating zone technique~\cite{Prabhakaran1} and subsequently characterized by powder X-ray diffraction and magnetic susceptibility measurements to confirm phase purity and magnetic ordering temperatures. Magnetic susceptibility measurements reveal two distinct anomalies around $T_\mathrm{N1} = 213$ K and $T_\mathrm{N2} = 230$ K, corresponding to successive magnetic phase transitions (Fig.~\ref{fig:S3}). The low-temperature magnetic structure ($T<T_\mathrm{N1}$) exhibits collinear antiferromagnetic ordering with propagation vector $(\nicefrac{1}{2},\ 0,\ {-\nicefrac{1}{2}})$, while the intermediate temperature range ($T_\mathrm{N1}<T<T_\mathrm{N2}$) is characterized by non-collinear spin-spiral ordering with propagation vector $(0.506,\ 0,\ -0.483)$~\cite{Aïn1}. As visualized through polarized optical microscopy (Fig.~\ref{fig:S4}), the grown crystal contains a minimal volume of twin domains, consistent with previous reports~\cite{Bush1}. Contributions from different domains are negligible in our time-resolved X-ray experiments because diffraction selects the properly oriented domain and diffuse scatterings appear in the vicinity of Bragg reflections.  Figures~\ref{fig:S5}\textbf{A} and \ref{fig:S5}\textbf{B} show a diffraction profile and a rocking curve around the $(\nicefrac{1}{2}\ 0\ {-\nicefrac{1}{2}})$ magnetic Bragg reflection measured at 929.9 eV, demonstrating the well-defined magnetic ordering and sample quality, respectively. Magnetic scattering produces significant linear polarization dependence. 

\subsubsection*{\underline{DFT calculations}}

\begin{figure}[ht]
    \centering
    \includegraphics[width=0.9\textwidth]{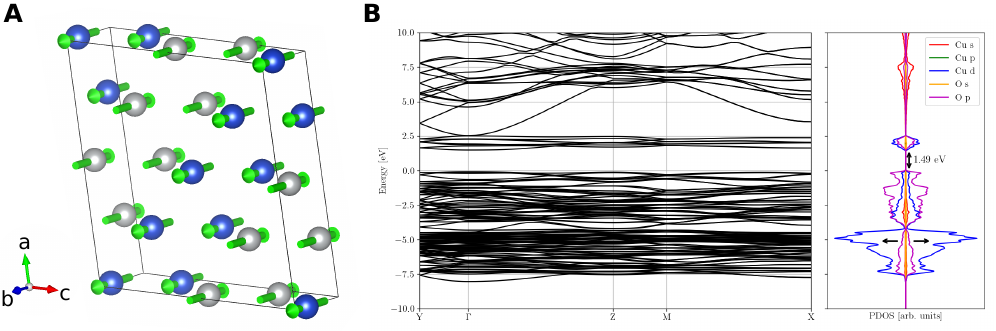}
    \caption{\textbf{Electronic structure of CuO}. (\textbf{A}) Antiferromagnetic unit cell of CuO, rotated to align the spin direction with the $z$ direction. The magnetic moments are indicated by arrows. (\textbf{B}) Band structure and corresponding partial density of states (PDOS).}
    \label{fig:band_structure}
\end{figure}

First-principles calculations were performed within density functional theory (DFT) using the Quantum ESPRESSO package~\cite{Giannozzi1}. The exchange–correlation functional was treated within the generalized gradient approximation using the Perdew–Burke–Ernzerhof (PBE) functional~\cite{Perdew1}. To account for strong electronic correlations in the Cu 3$\textit{d}$ orbitals, we employed the DFT+U method with an on-site Hubbard interaction \textit{U}=7.0 eV applied to the Cu 3$\textit{d}$ states~\cite{Ekuma1}. 
Antiferromagnetic (AFM) order was imposed in accordance with experimentally established magnetic structures (Fig.~\ref{fig:band_structure}\textbf{A}). For convenience, the unit cell was rotated such that the crystallographic $b$ direction aligns with the spin quantization ($z$) axis. Within this collinear AFM configuration, we obtained an insulating ground state with a band gap (Fig.~\ref{fig:band_structure}\textbf{B}) of 1.49 eV and a Cu magnetic moment of  $\mu=0.61 \mu_\text{B}$, in good agreement with experimental values~\cite{Wang1,Nolan1,Ghijsen1} and previous theoretical studies~\cite{Wu2, Heinemann1, Ekuma1}.

\subsubsection*{\underline{Wannier Construction and Time-Dependent Simulations}}

Following the convergence of the ground-state electronic structure, we constructed a Wannier Hamiltonian using the Wannier90 package~\cite{Pizzi1}. The Wannierization process included Cu-3$\textit{d}$ and O-2$\textit{p}$ orbitals to accurately capture the relevant states. Time-dependent simulations were then performed using the dynamics-w90 code~\cite{Michaelschueler1}, incorporating an external electric field pulse of the form
\begin{align}
\vec{E}(t) = \vec{E}_0 e^{-(t-t_0)^2/2T^2_p} \sin(\omega (t-t_0)) \ ,
\end{align}
with a pulse duration $T_p$ corresponding to a full width at half maximum (FWHM) of 35 fs and a frequency $\omega = 3.1$ eV, consistent with experimental conditions. The density matrix was propagated in time according to the equation of motion
\begin{align}
i \frac{d}{dt} \rho(\vec{k}, t) = [H(\vec{k}, t), \rho(\vec{k}, t)] \ .
\end{align}
Combining the occupations after the pump with the orbital-resolved density of states then yields the plot in Fig.~\ref{fig:2}\textbf{B}.

\subsubsection*{\underline{Spin Model and Magnon Dispersion}}

To verify that the effective magnetic interactions underlying the calculations reproduce the experimentally observed spin excitations, we analyzed a simplified spin model derived from the electronic structure. The Cu$^{2+}$ions in CuO carry localized spin-1/2 moments, which can be described by an effective Heisenberg Hamiltonian obtained from the strong-coupling limit of the Hubbard model.

For this purpose, we adopted a simplified orthorhombic unit cell containing one Cu site per layer, neglecting the small monoclinic distortion and the weak interlayer structural complexity (Fig.~\ref{fig:spin_model}\textbf{A}). This approximation captures the dominant in-plane magnetic interactions relevant to the low-energy magnon spectrum. The model includes nearest- and next-nearest-neighbor exchange interactions with representative values $J_\mathrm{AFM} = 120$~meV, $J_\mathrm{FM} = -15$~meV, and an interlayer coupling of $J_\mathrm{out}=-18$~meV, as well as a small easy-axis anisotropy $D=0.025$~meV.

The magnon dispersion was calculated within linear spin-wave theory. The resulting acoustic magnon branch (Fig.~\ref{fig:spin_model}\textbf{B}) closely reproduces the experimentally measured dispersion obtained from inelastic neutron scattering~\cite{Aïn1}, validating both the simplified orthorhombic description and the effective magnetic parameters used in the analysis.

\subsubsection*{\underline{Magnon excitation and magnon scattering}}

This feedback of charge dynamics onto the spin sector was incorporated by mapping the time-dependent electronic state onto an effective Hubbard model in the Mott-insulator regime. Following Ref.~\cite{Huang1}, in the limit of large $U$ and weak excitation, the Hubbard model can be approximated by spin excitations coupled to doublons (corresponding to Cu sites with $d^{10}$ configuration) and holons ($d^8$ configuration). The creation of spin waves due to the presence of photoexcitation doublons is then modeled by the sudden ramp of the doublon number and the associated spin-charge coupling. The excitation parameters were chosen to be consistent with the experiment. Details are described in the Supplementary Text.

The photoexcited magnon distribution is then taken as initial state for the magnon thermalization dynamics. The magnon-magnon scattering is treated on the level of the quantum Boltzmann equation (QBE), as described in the main text. The scattering integral in Eq.~\eqref{eq:QBE} is computed on a $N = 32\times 48 \times 32$ grid of k-points in the magnetic Brillouin zone, similar to electron-electron scattering. The magnon-magnon interaction vertex is derived from the effective Heisenberg Hamiltonian by including the leading order beyond linear spin wave theory (see Supplementary Text for details).
Similarly, magnon-phonon scattering is modeled by parameterizing the relevant phonon modes known from experiments~\cite{Reichardt1} and deriving the magnon-phonon coupling from the Heisenberg Hamiltonian. This coupling is then used to define the corresponding QBE scattering integral.

% The Materials and Methods section should contain details of the samples measured,
% experiments performed, observations taken, simulations run, data analysis, statistical methods etc.
% Give enough detail for any competent researcher in your field to fully reproduce the results.

% To refer to this section from the main text, use the numbered note in the reference list \cite{methods}.
% Refer to figures and tables in the same way as in the main text but now all capitalized e.g.
% Fig.~\ref{fig:example}, Table~\ref{tab:example},
% Fig.~\ref{fig:sup_example} and Table~\ref{tab:sup_example}.
% Cite references in the usual way \cite{example2},
% including any that are only cited in the supplement \cite{sm_example,conference_example}.

% The numbering of figures, tables, equations and pages has been reset to start from S1, as in
% \begin{equation}
% 	\cos(2\theta) = \cos^2\theta - \sin^2\theta.
% 	\label{eq:sup_example} % Use a logical label
% \end{equation}

% \subsubsection*{Example supplement heading}

% The two main sections of the supplement can be split up using headings.

%%%%%%%%%%%%%%%% SUPPLEMENTARY TEXT %%%%%%%%%%%%%%%
\newpage
\subsection*{Supplementary Text}
% The Supplementary Text section can only be used to directly support statements made in the main text
% e.g. to present more detailed justifications of assumptions, investigate alternative scenarios,
% provide extended acknowledgements etc.
% Material in this section cannot claim results or conclusions that weren't mentioned in the main text.
% To refer to this section from the main text, just write (Supplementary Text).

\subsubsection*{\underline{Deconvolution of RXD and RDS signals}}

RDS signals exhibit enhanced intensity in the vicinity of the $(\nicefrac{1}{2}\ 0\ {-\nicefrac{1}{2}})$ magnetic Bragg reflection due to the energy minimum in the magnon dispersion, leading to a large population of the bosonic excitations. Resonant X-ray scattering measurements inherently contain contributions from both RXD and RDS components, which show distinct dynamics following photoexcitation. While ultrafast demagnetization produces a quasi-instantaneous decrease in RXD intensity, the subsequent creation and redistribution of magnon density manifest as an increase in RDS intensity at later timescales. To capture these dynamics, we analyzed data using two exponential functions: 
\begin{equation}
	I_{\mathrm{ON}}(t) - I_{\mathrm{OFF}} 
    = A_{\mathrm{RXD}} \exp\left(-t/\tau_{\mathrm{RXD}}\right) 
    + A_{\mathrm{RDS}} \exp\left(-t/\tau_{\mathrm{RDS}}\right).
	\label{eq:S1}
\end{equation}

\begin{figure} 
	\centering
	\includegraphics[width=\textwidth]{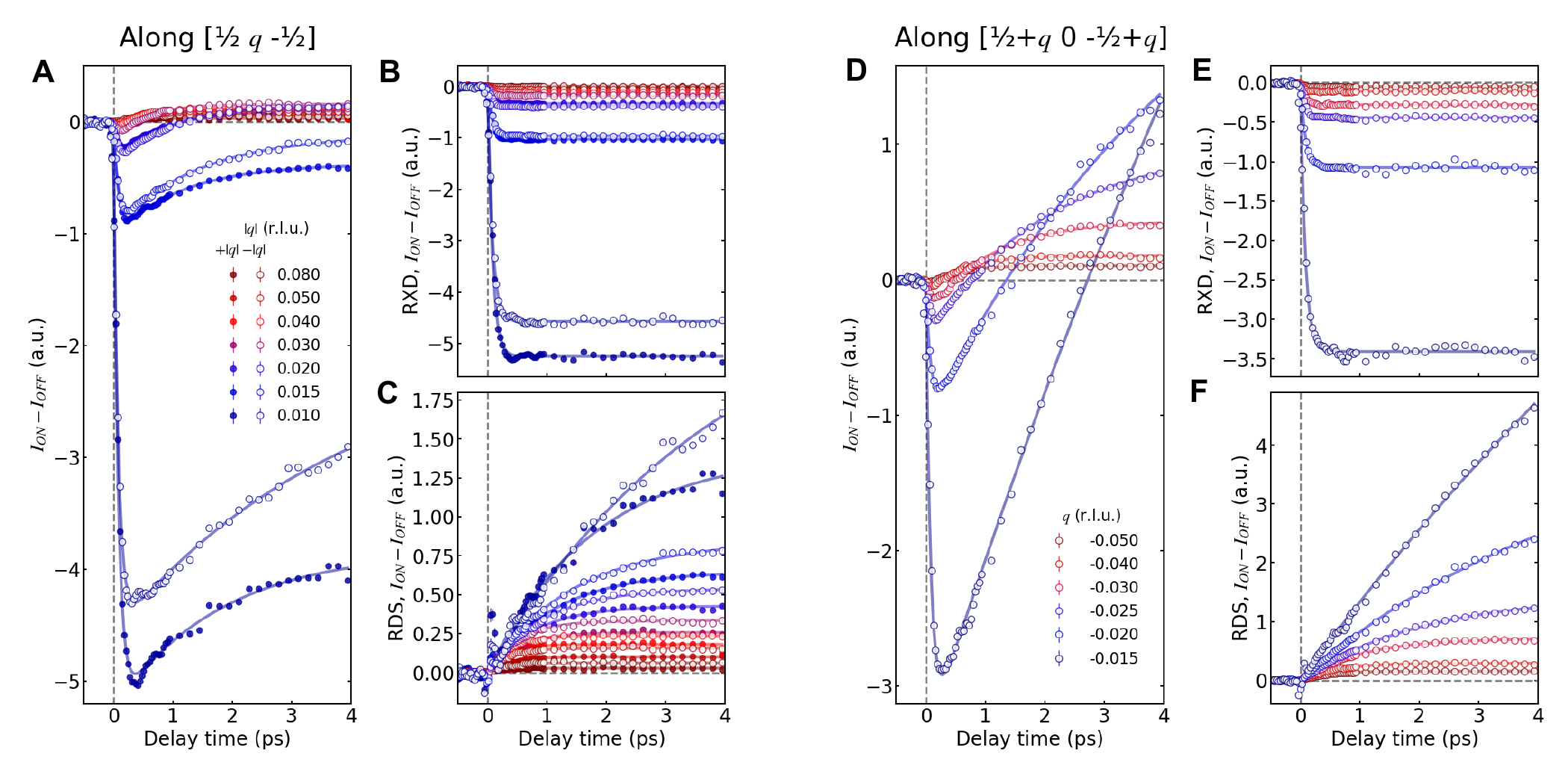}
	\caption{\textbf{Deconvolution of time-resolved resonant X-ray scattering intensities into RXD and RDS components.}
        \textbf{A} is identical as Fig.~\ref{fig:3}\textbf{A}. Momentum dependence of time-dependent resonant X-ray scattering intensities around the D point (see main text) (\textbf{A} and \textbf{D}) was deconvoluted into two exponential terms representing the RXD (\textbf{B} and \textbf{E}) and RDS (\textbf{C} and \textbf{F}) components. Solid curves represent fits using two exponential terms to separate the RXD drop and RDS upturn components. Filled (open) points indicate data measured at positive (negative) $q$.
		}
	\label{fig:S6}
\end{figure}

\noindent
Here, $A_{\mathrm{RXD}}$ ($< 0$) and $A_{\mathrm{RDS}}$ ($> 0$) represent the amplitude, and $\tau_{RXD}$ and $\tau_{RDS}$ denote the time constant for the photo-induced changes in RXD and RDS intensities, respectively. The deconvolution of the momentum-dependent data presented in Fig.~\ref{fig:3}\textbf{A} into the RXD and RDS components is shown in Figs.~\ref{fig:S6}\textbf{A}--\ref{fig:S6}\textbf{C} using this model.  

\subsubsection*{\underline{Transient incommensurate state at high excitation regime}}
As shown in the main text, moderate excitation fluence leads to strongly non-thermal magnons but keeps the long-range collinear AFM state because the average temperature remains below $T_\mathrm{N1}$. Although a photo-induced phase transition into the incommensurate phase~\cite{Johnson1} could result in an upturn signal in a certain momentum range via the transient appearance of a corresponding magnetic Bragg peak, it does not explain our experimental observation in the moderate fluence regime. This is because (i) the measured momentum point $(0.49, 0.01, -0.51)$ is distant from the incommensurate peak position at $(0.506, 0, -0.483)$, (ii) it does not explain the systematic tail intensity increase around the magnetic Bragg peak (Fig.~\ref{fig:2}\textbf{C}) and (iii) the momentum-dependent time constant (Fig.~\ref{fig:3}\textbf{F}), and (iv) we did not experimentally detect any significant upturn signal at the incommensurate peak position compared to other momentum points after the photoexcitation within our experimental precision. The overall temperature increase estimated by the specific heat~\cite{Gmelin1} and deposited energy density is below 20 K ($<$ $T_\mathrm{N1}$), indicating the remaining collinear AFM state after the photoexcitation. 
%Using a larger excitation laser fluence can lead to the transient incommensurate state with photoexcited magnons and full suppression of the commensurate order (Fig.~\ref{fig:S7}) once the average temperature in the probed volume goes above $T_\mathrm{N1}$.

\begin{figure} 
	\centering
	\includegraphics[width=0.8\textwidth]{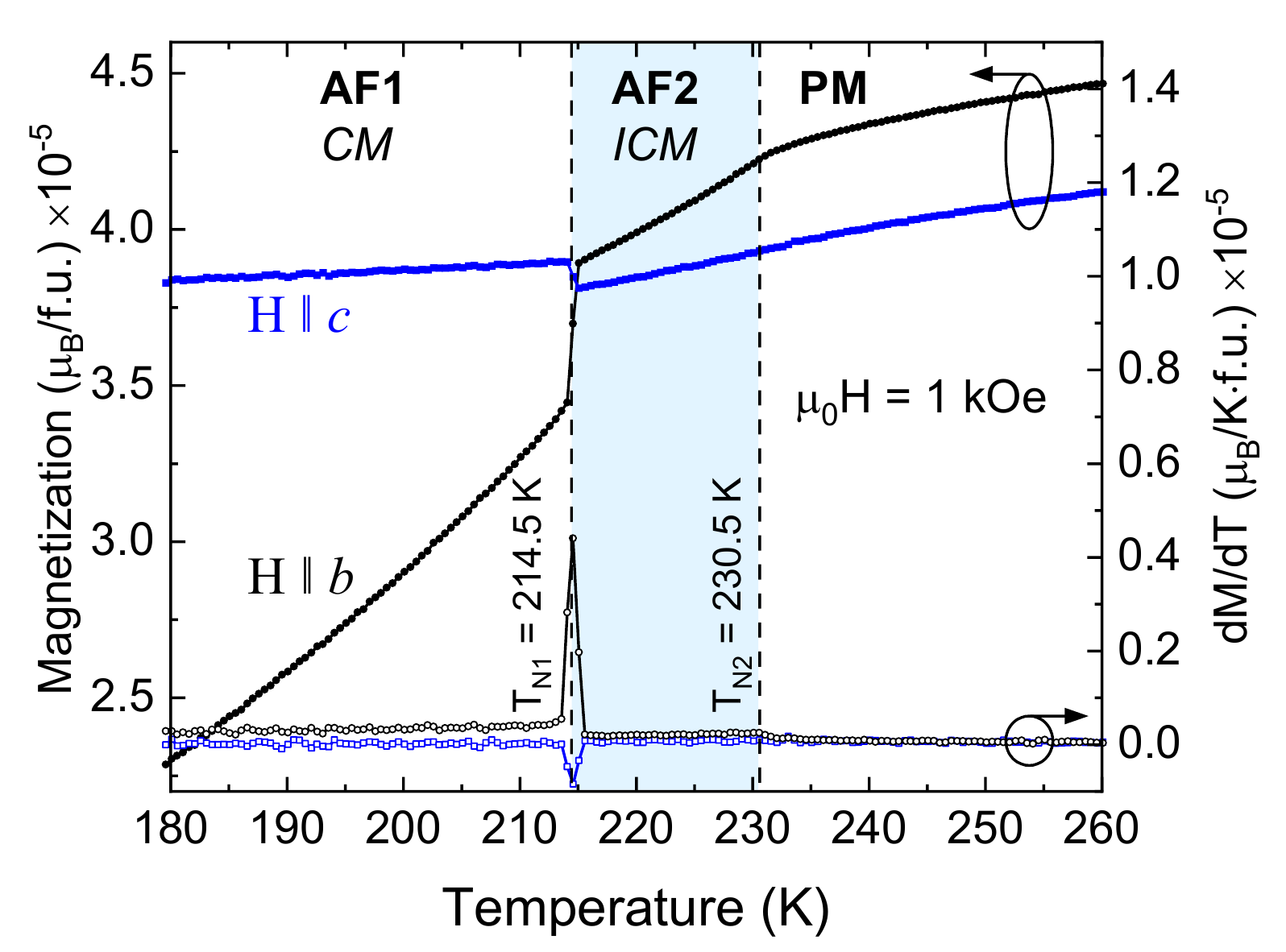}
	\caption{\textbf{Temperature dependence of magnetic susceptibility of the CuO single crystal sample.}
        ICM and CM represent the incommensurate magnetic phase and commensurate magnetic phase, respectively. 
		}
	\label{fig:S3}
\end{figure}

\begin{figure} 
	\centering
	\includegraphics[width=0.6\textwidth]{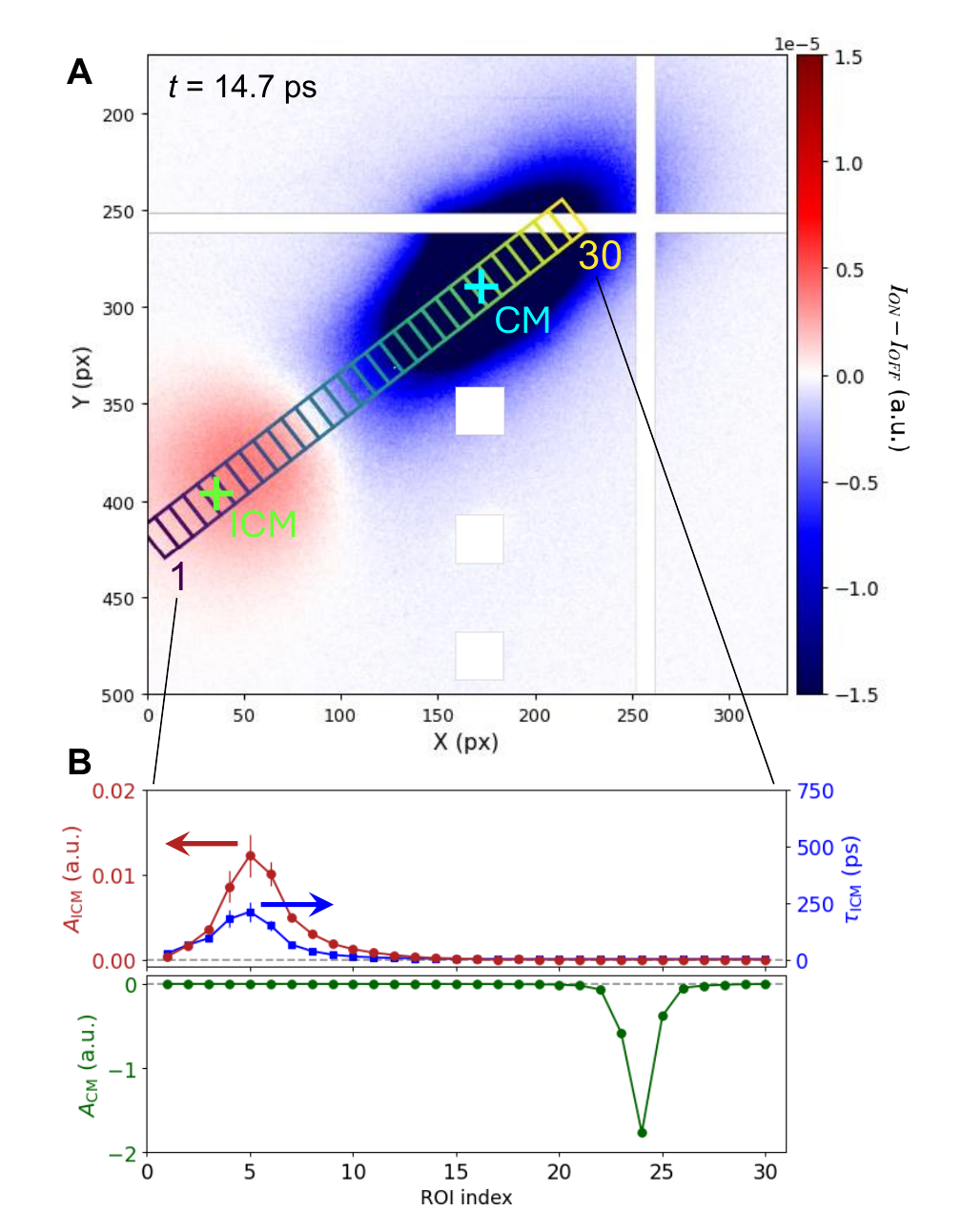}
	\caption{\textbf{Transient appearance of incommensurate order at high excitation regime.}
        (\textbf{A}) Difference detector image ($I_{\mathrm{ON}}-I_{\mathrm{OFF}}$) of resonant X-ray scattering at a pump-probe delay of $t$ = 14.7 ps. The blue (negative) region corresponds to the suppressed CM Bragg peak $(\nicefrac{1}{2}\ 0\ {-\nicefrac{1}{2}})$; the red (positive) region indicates the transient appearance of the ICM peak. Colored rectangles mark 30 regions of interest (ROIs) along the direction connecting the ICM and CM peak positions; white areas are masked detector defects. (\textbf{B}) Fit parameters extracted for each ROI by fitting the time-dependent signal with a double-exponential model analogous to Eq. \ref{eq:S1}. Top: amplitude $A_{\mathrm{ICM}}$ (red, left axis) and time constant $\tau_{\mathrm{ICM}}$ (blue, right axis) of the component associated with the growth of the ICM order. Bottom: amplitude $A_{\mathrm{CM}}$ (green) of the CM peak suppression component. The two distinct dynamics develop symmetrically about the CM and the emerging ICM positions in reciprocal space.
		}
	\label{fig:S7}
\end{figure}

Increasing the excitation fluence raises the average temperature above $T_\mathrm{N1}$ and, as expected from the equilibrium phase diagram (Fig.~\ref{fig:S3}), stabilizes the incommensurate spin-spiral phase, consistent with the previous study\cite{Johnson1}. Figure~\ref{fig:S7}\textbf{A} shows a two-dimensional RXD profile at 14.7 ps after the photoexcitation. While the region with an intensity decrease corresponds to the commensurate peak, the region with an intensity increase corresponds to the incommensurate peak that appears transiently. Figure~\ref{fig:S7}\textbf{B} shows the upturn (top row) and reduction (bottom row) amplitudes and time constants of integrated time traces in multiple regions of interests along the direction connecting the two peak positions. While the reduction amplitude is symmetric around the commensurate peak, the upturn amplitude and time constant are symmetric around the incommensurate peak, which is in contrast to the tr-RDS signals symmetric around $\textbf{q}_{\text{RDS}}$ (see main text). The larger time constant of the incommensurate peak time traces than the previous study~\cite{Johnson1} is due to the absence of the coexistence of the two orders, which may appear in a specific temperature window. The incommensurate order can respond rapidly by moving domain walls when the two phases coexist. 
%show a delay in the appearance of the incommensurate peak, compared to the quasi-instantaneous intensity reduction of the commensurate peak. The retardation is in contrast to the previous study~\cite{Johnson1}, which investigated the dynamics of the two orders coexisting within a specific temperature window, enabling the incommensurate order to respond rapidly by moving domain walls. 
%The slow time evolution of the incommensurate peak intensities indicates the timescale for exchange interactions to wind the spins into a long-range spiral. 
%Both amplitude and time constant of the upturn signal are symmetric around the incommensurate peak position, in contrast to tr-RDS signals symmetric around $\textbf{q}_{\text{RDS}}$. 

\subsubsection*{\underline{Additional evidence of magnon density redistribution}}

\begin{figure} 
	\centering
	\includegraphics[width=0.8\textwidth]{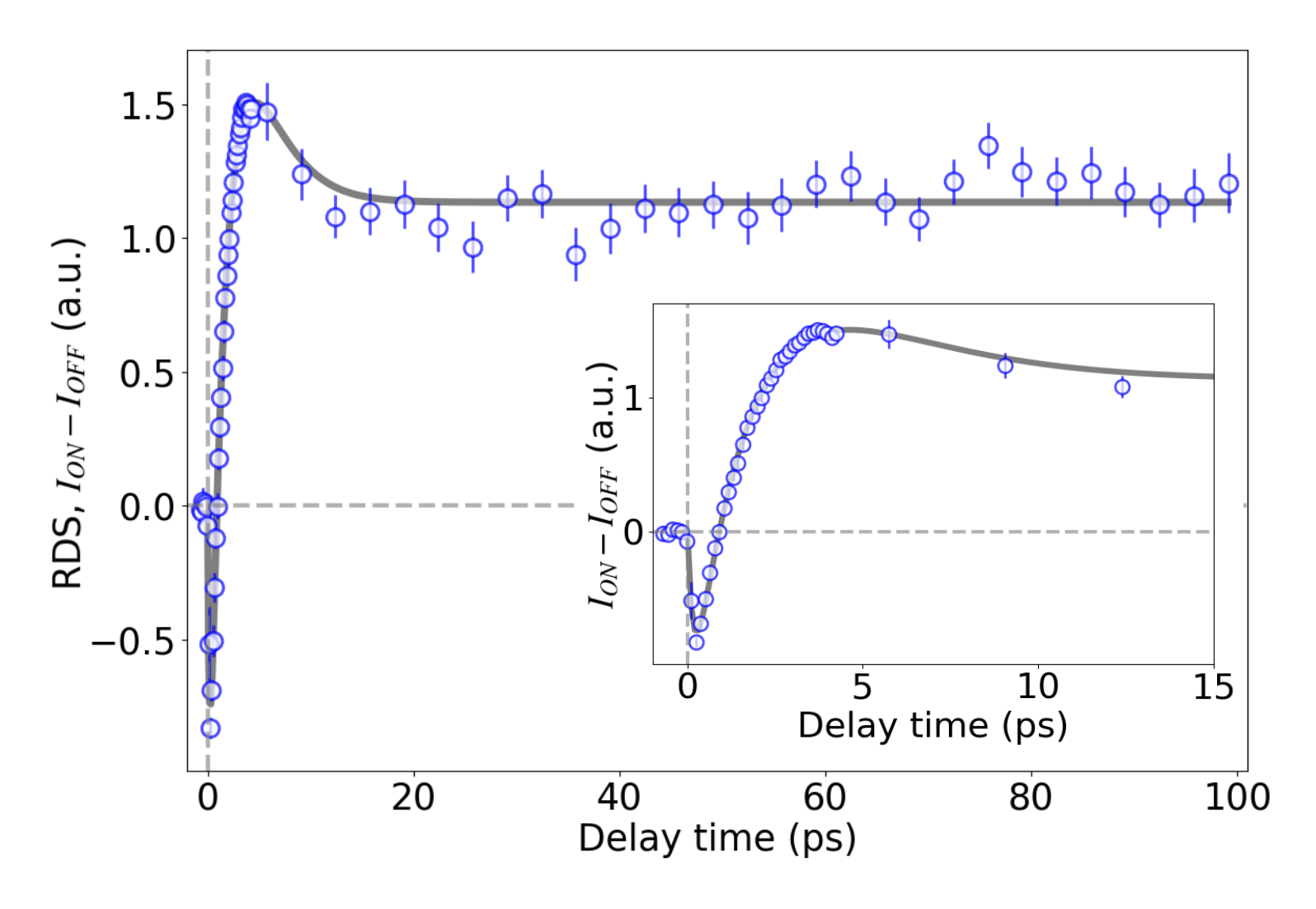}
	\caption{\textbf{Time trace of RDS at (0.48, 0.01, $\mathbf{-}$0.52).}
        Errors are standard errors.
		}
	\label{fig:S8}
\end{figure}

RDS signals can exhibit more complex time evolution than the simple phenomenological model with two exponential functions due to the magnon density redistribution processes that depend on both timescale and momentum point. Figure~\ref{fig:S8} represents a time trace of resonant X-ray scattering intensity measured at $(0.48,\ 0.01,\ -0.52)$. The initial drop is due to ultrafast demagnetization causing RXD intensity reduction, followed by clear RDS upturn signals. However, the subsequent evolution is non-monotonic and exhibits a maximum around 5 ps and plateaus after $\sim$10 ps until further relaxation occurs via the anomalous scattering. These intricate dynamics further support the interpretation of magnon density redistribution within the acoustic magnon branch, in addition to possible contributions from optical magnon decay and the magnon-number conserved scattering (main text).

\subsubsection*{\underline{Momentum dependence along $[h\ 0\ h]$}}

\begin{figure} 
	\centering
	\includegraphics[width=0.8\textwidth]{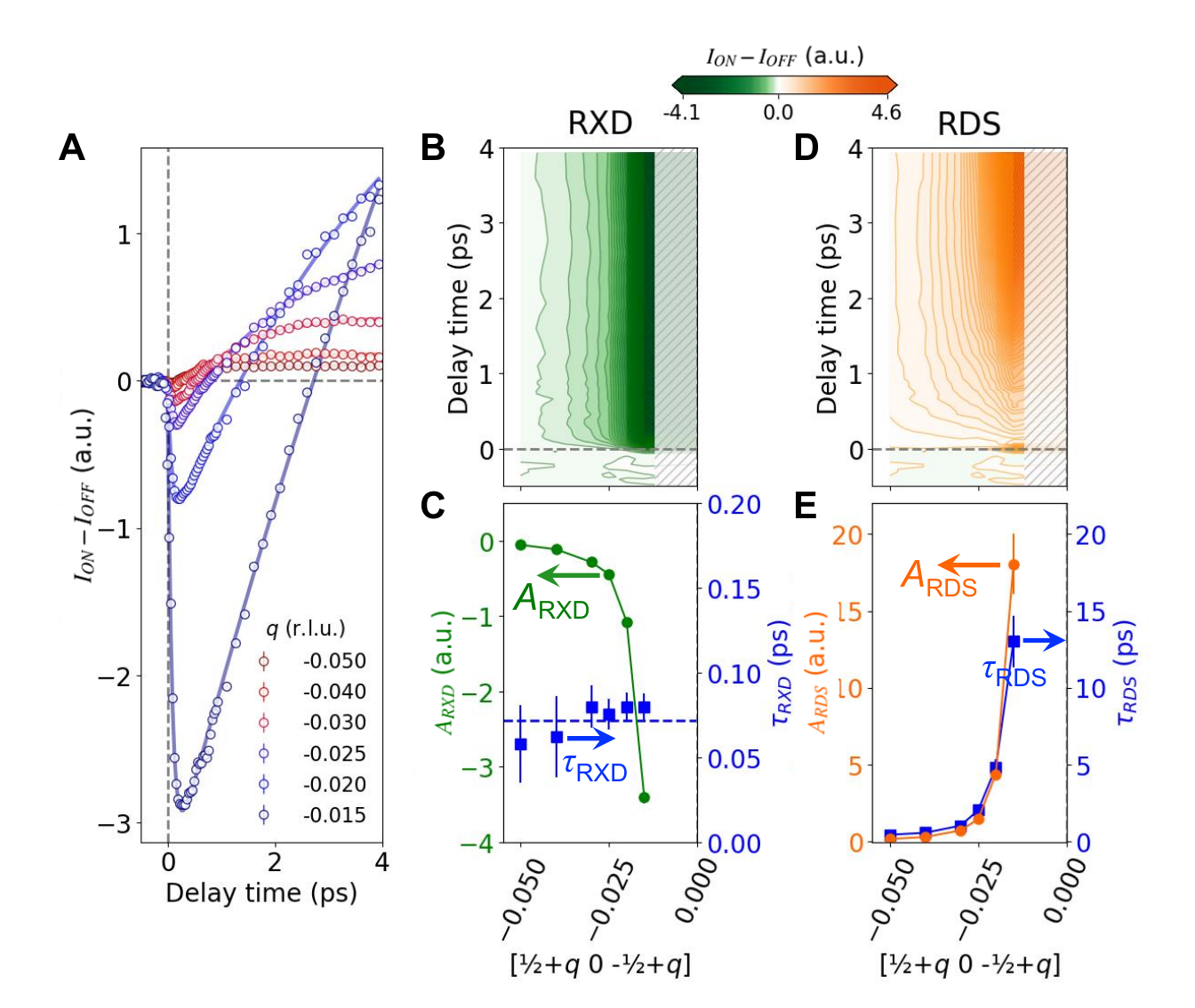}
	\caption{\textbf{Momentum dependence of time-resolved resonant diffuse scattering intensities along [\textit{h} 0 \textit{h}].}
        (\textbf{A}) Time traces of RDS signals at momentum points around the D point (see main text) along $[h\ 0\ h]$. Solid curves represent fits using two exponential terms to separate the RXD drop and RDS upturn components. (\textbf{B}, \textbf{D}) Momentum-delay time contour plots of the RXD (\textbf{B}) and RDS (\textbf{D}) components derived from the time traces in \textbf{A}. (\textbf{C}, \textbf{E}) Momentum dependence of RXD-drop (\textbf{C}) and RDS-upturn (\textbf{E}) amplitudes and their time constants extracted from fits to the data in \textbf{A}. Error bars represent standard error in \textbf{A} and are from the fit in \textbf{C} and \textbf{E}. 
		}
	\label{fig:S9}
\end{figure}

We measured the momentum dependence of tr-RDS time traces along $[h\ 0\ h]$ (Fig. S9), in addition to along $\mathbf{b}^\ast$ (Fig.~\ref{fig:3}). The same fitting procedure using Eq.~\ref{eq:S1} allows us to decompose the time traces into RXD (Fig.~\ref{fig:S9}\textbf{B}) and RDS (Fig.~\ref{fig:S9}\textbf{D}) components. While the RXD component shows a similar amplitude and the same time constant within the error bars as the dataset along $\mathbf{b}^\ast$, the RDS component shows substantially larger upturn signals. This indicates strong anisotropy in the monoclinic structure. 

\subsubsection*{\underline{"Temperature" estimation}}

The non-thermally distributed magnon densities allow us to estimate corresponding "temperatures" as a function of momentum. When an excitation predominantly couples to a specific mode, the phenomenological temperature model requires subdividing the subsystem: for example, hot optical phonons in the lattice\cite{Perfetti1}. In our case, the spin system is subdivided into hot acoustic magnons. Hot-magnon temperatures at fast timescales reach $\sim$500 K at $(0.5, 0.1, -0.5)$, far exceeding the bulk quasi-equilibrium limit of $\sim$210 K. This indicates that the excitation energy initially deposited in the electronic system, resulting in doublon creation that quickly decays, transfers into the magnonic subsystem at ultrafast timescales, while the electronic states return to equilibrium~\cite{Novelli1, Baykusheva1} and the average temperature, except for hot carriers, remains below $T_\mathrm{N1}$.

\subsubsection*{\underline{Tr-NRDS of phonon origin}}

\begin{figure} 
	\centering
	\includegraphics[width=\textwidth]{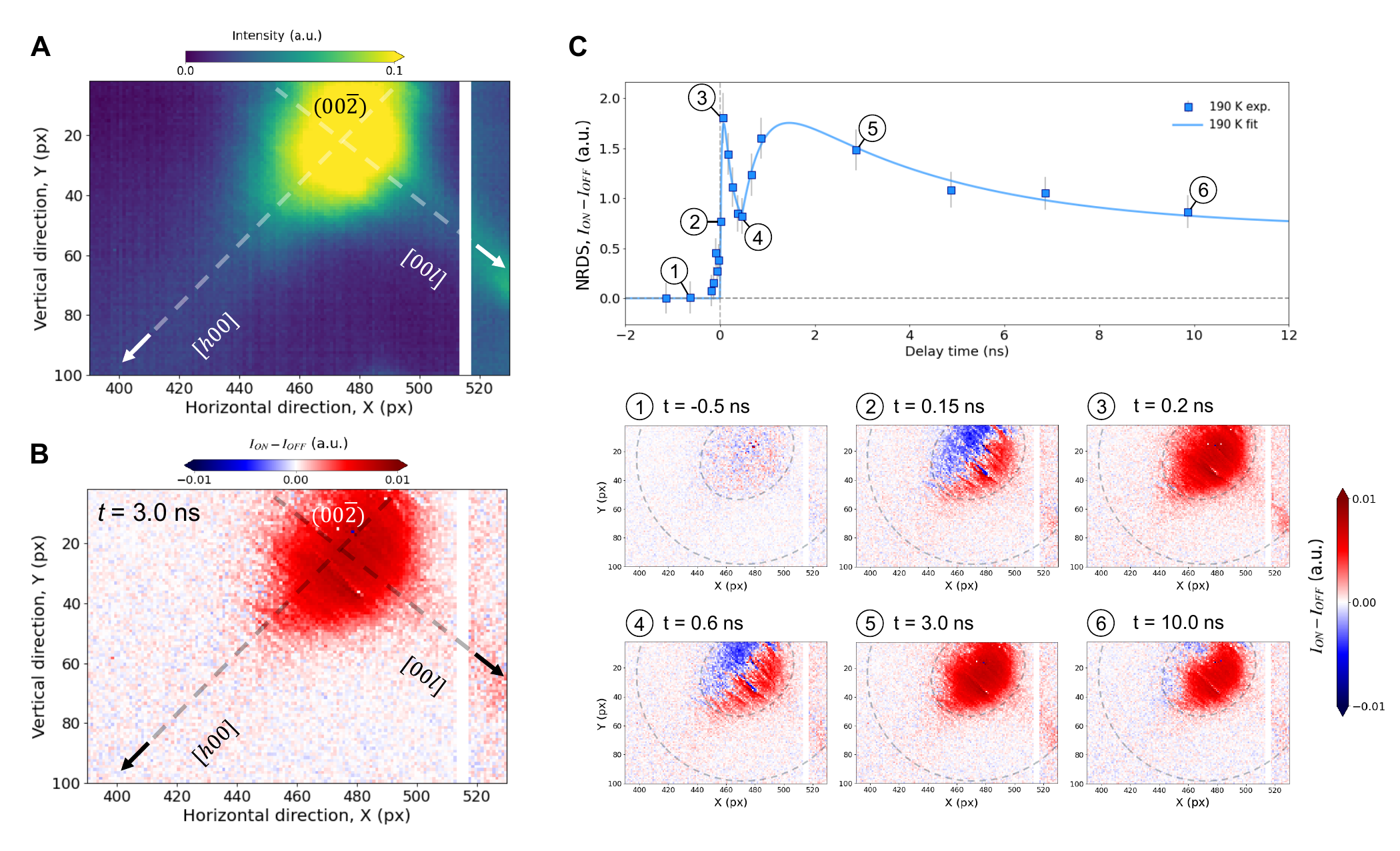}
	\caption{\textbf{Tr-NRDS signals on a two-dimensional detector.}
        ($\mathbf{A}$) (0 0 $-$2) diffraction pattern in equilibrium. The color scale is saturated at the high-intensity side to enhance the visibility of weak features. Streaks along $[0\ 0\ l]$ and $[h\ 0\ 0]$ indicate NRDS by pre-existing phonons due to thermal energy and static disorder. ($\mathbf{B}$) Pump-induced change in X-ray scattering intensities at 3 ns after the photoexcitation. ($\mathbf{C}$) Time trace of tr-NRDS intensities, the same as the blue plot in Fig.~\ref{fig:4}\textbf{D}. Bottom panels display the pump-induced change in two-dimensional X-ray scattering intensities at respective delay times indicated by the numbers in the top panel. 
		}
	\label{fig:S10}
\end{figure}

As described in the main text, photoexcited magnon energy is transferred to phonons via the anomalous scattering and subsequently dissipated into the lattice, resulting in possible thermal expansion and corresponding Bragg peak shifts. Figures~\ref{fig:S10}\textbf{A} and \ref{fig:S10}\textbf{B} display the equilibrium two-dimensional X-ray scattering intensity pattern around the $(0\ 0\ {-2})$ Bragg reflection and pump-induced change in the pattern at 3 ns after the photoexcitation, representing a tr-NRDS intensity profile. Phonon density transiently increases along $[0\ 0\ l]$ and $[h\ 0\ 0]$ on top of pre-existing phonons in equilibrium, as marked by streaks, due to less steep acoustic phonon dispersions along these directions compared to other directions~\cite{Reichardt1}. We chose the region of interest between two dashed ellipses with different sizes as shown in Fig.~\ref{fig:S10}\textbf{C} to avoid possible contamination by Bragg peak shift, which is clear along $[0\ 0\ l]$ at delay times where tr-NRDS signals are less substantial, e.g., at 0.15 ns and 0.6 ns. Consequently, we can conclude that our tr-NRDS time traces are due to diffuse scattering by phonons. 

\subsubsection*{\underline{Phonon dissipation process}}

\begin{figure} 
	\centering
	\includegraphics[width=0.6\textwidth]{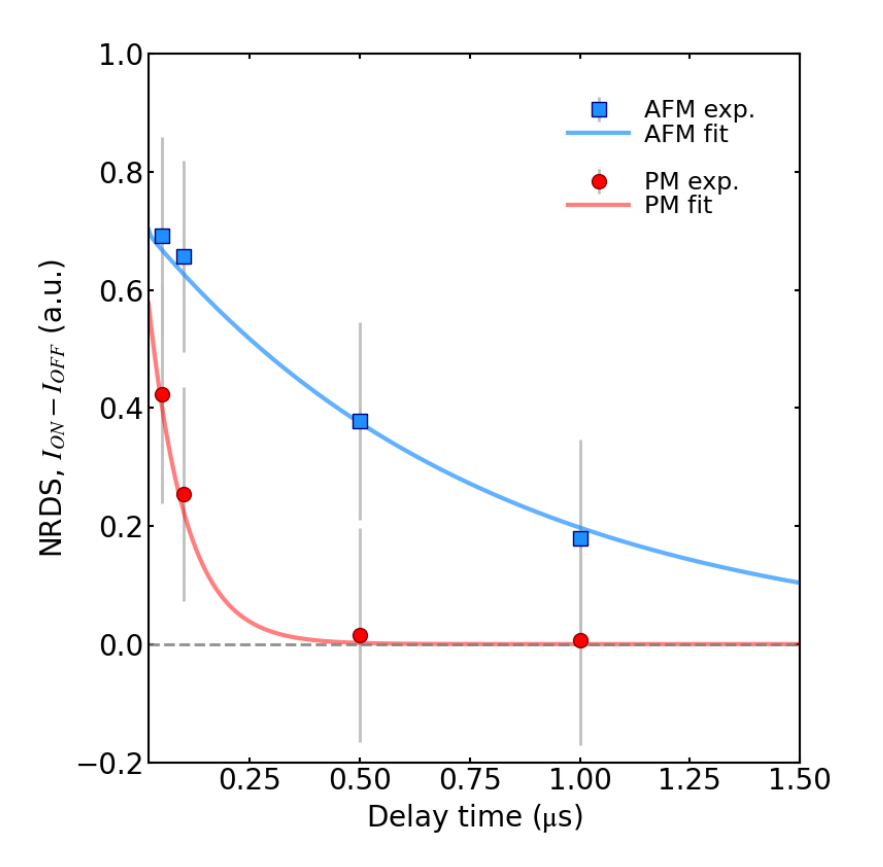}
	\caption{\textbf{Long timescale comparison of tr-NRDS intensities between the AFM phase and PM phase.}
		}
	\label{fig:S11}
\end{figure}

Figure~\ref{fig:S11} shows a long timescale comparison of tr-NRDS intensities between the AFM phase (190 K) and the PM phase (300 K), demonstrating the relaxation process of phonons at a microsecond timescale. While the time constant is $\sim$0.1 $\upmu$s in the PM phase, it is $\sim$0.7 $\upmu$s in the AFM phase. Although the slow energy transfer process from magnons via the anomalous scattering creates additional phonons in the AFM phase, the phonon diffusion efficiency should be substantially higher at low temperatures due to a larger phonon mean path length, resulting in a larger phonon thermal conductivity\cite{Akopyan1}. The temperature dependence is not explained by the previously proposed magnon scenario, either~\cite{Akopyan1}. Thus, it is indicated that the electronic thermal conductivity significantly contributes to the energy dissipation process, likely due to oxygen excess\cite{Zheng1} and Cu vacancies\cite{Wu1}. 

\subsubsection*{\underline{Theory of spin-charge coupling in the photoexcited state}}

The $d^9$ configuration of the Cu$^{2+}$ ions in CuO gives rise to a localized spin-1/2 moment on each Cu site. This situation can be described by the Hubbard model at half-filling in the limit of strong on-site Coulomb repulsion $U$:
\begin{align}
    \label{eq:Hubbard}
    \hat{H} = -t_0 \sum_{\langle\vec{R}, \vec{R}^\prime \rangle , \sigma}  \hat{c}_{\vec{R}, \sigma}^\dagger \hat{c}_{\vec{R}^\prime, \sigma}  + U \sum_{\vec{R}} \hat{n}_{\vec{R}, \uparrow} \hat{n}_{\vec{R}, \downarrow} \ .
\end{align}
Here, $\hat{c}_{\vec{R}, \sigma}^\dagger$ ($\hat{c}_{\vec{R}, \sigma}$) creates (annihilates) an electron with spin $\sigma$ on site $\vec{R}$, and $\hat{n}_{\vec{R}, \sigma} = \hat{c}_{\vec{R}, \sigma}^\dagger \hat{c}_{\vec{R}, \sigma}$ is the corresponding number 
operator.

In the language of the Hubbard model~\eqref{eq:Hubbard}, the photodoping excitation can be described as the creation of additional doublons (doubly occupied Cu sites). Previous studies on the Hubbard model have already demonstrated the strong impact of photodoping on the AFM order~\cite{Gillmeister1}. Furthermore, as shown in Ref.~\cite{Golež1}, the Coulomb interaction between the photodoped carriers in the ligand orbitals and the upper Hubbard band provides another channel of ultrafast doublon-holon creation. This mechanism is the basis for the ultrafast suppression of AFM order observed in our experiments. Additionally, the O-$2p$ holes can create holons via the coupling to the lower Hubbard band, creating an another source of the supression of AFM order. 

When the system is photoexcited, additional doublons (doubly occupied Cu sites) are created. These photodoped carriers can interact with the localized spins, leading to a modification of the exchange interaction and consequently affecting the AFM order. The presence of doublons and holons introduces new channels for spin fluctuations and can enhance spin-charge coupling in the system.

To describe the effect of photodoping on the spin dynamics, we follow the approach from Ref.~\cite{Huang1}. In this framework, the Hubbard model~\eqref{eq:Hubbard} is mapped onto an effective Hamiltonian including doublon, holons, and spin degrees of freedom:
\begin{align}
    \label{eq:Heff}
    \hat{H}_\mathrm{eff} = -B_0 \sum_{\langle \vec{R}, \vec{R}^\prime \rangle} \left( \hat{d}_{\vec{R}}^\dagger 
    \hat{d}_{\vec{R}^\prime} -  \hat{e}_{\vec{R}}^\dagger 
    \hat{e}_{\vec{R}^\prime} \right) \left(\hat{s}_{\vec{R}} + \hat{s}^\dagger_{\vec{R}^\prime} \right) + \hat{H}_\mathrm{spin} \ .
\end{align}
Here, $B_0 = \sqrt{(m + 1/2)} t_0$, where $m$ is magnetization. The operators $\hat{d}_{\vec{R}}^\dagger$ ($\hat{d}_{\vec{R}}$) and $\hat{e}_{\vec{R}}^\dagger$ ($\hat{e}_{\vec{R}}$) create (annihilate) doublons and holons on site $\vec{R}$, respectively. The effective bosonic annihilation operator $\hat{s}_{\vec{R}}$ acts on the localized spin-1/2 moment at site $\vec{R}$. For the pure Hubbard model,  the spin Hamiltonian $\hat{H}_\mathrm{spin}$, up to second order in $t_0/U$, is given by:
\begin{align}
    \label{eq:Hspin}
    \hat{H}_\mathrm{spin} = 4(2m  + 1) J \sum_{\vec{R}} \hat{s}^\dagger_{\vec{R}} \hat{s}_{\vec{R}} + \left(m + \frac12\right) J  \sum_{\langle \vec{R}, \vec{R}^\prime \rangle} \left(\hat{s}^\dagger_{\vec{R}} \hat{s}^\dagger_{\vec{R}^\prime} + \hat{s}_{\vec{R}} \hat{s}_{\vec{R}^\prime} \right) \ ,
\end{align}
where $J = 4 t_0^2/U$ is the exchange interaction between neighboring spins. The spin Hamiltonian~\eqref{eq:Hspin} describes the magnon excitations in the AFM state, and it can be paralleled with the linear spin wave theory for the Heisenberg model. 

To simplify the following calculations, we consider a simplified unit cell containing only one Cu site by ignoring the atoms between the layers along the $c$ direction (Fig.~\ref{fig:spin_model}\textbf{A}). We further ignore the small monoclinic distortion and treat the system as orthorhombic. Introducing next-nearest-neighbor exchange couplings $J_\mathrm{FM}=-15$ meV, $J_\mathrm{AFM} = 120$ meV, and $J_\mathrm{out} = -18$ meV (see Fig.~\ref{fig:spin_model}\textbf{B}), as well as a small easy-axis anisotropy $D=0.025$ meV, the magnon dispersion obtained from linear spin-wave theory (Fig.~\ref{fig:spin_model}\textbf{C}) closely resembles the experimentally measured dispersion of the acoustic magnon branch in CuO as determined by inelastic neutron scattering~\cite{Aïn1}. The values chosen here are approximately consistent with calculated exchange couplings~\cite{Giovannetti1}.

There are two inequivalent lattice sites in the AFM unit cell; denoting the corresponding bosonic operators $\hat{b}_{\vec{R}A} = \hat{s}_{\vec{R}}$ for site A and $\hat{b}_{\vec{R}B} = \hat{s}_{\vec{R}}$ for site B and introducing the Fourier transform $\hat{b}_{\vec{k}\alpha} = 1 / \sqrt{N} \sum_{\vec{R}} e^{-i \vec{k} \cdot \vec{R}} \hat{b}_{\vec{R}j}$ ($j = A, B$), the spin Hamiltonian~\eqref{eq:Hspin} can be written as
\begin{align}
    \label{eq:Hspin_2}
    \hat{H}_\mathrm{spin} = \sum_{\vec{k}} \vec{x}^\dagger_{\vec{k}} \vec{H}(\vec{k}) \vec{x}_{\vec{k}} \ .
\end{align}
Here, we introduced the vector $\vec{x}_{\vec{k}} = (\hat{b}_{\vec{k}A}, \hat{b}_{\vec{k}B}, \hat{b}^\dagger_{-\vec{k}A}, \hat{b}^\dagger_{-\vec{k}B})^T$ and the $4 \times 4$ matrix
\begin{align}
    \vec{H}(\vec{k}) = \begin{pmatrix}
    A(\vec{k}) - C & B_{\vec{k}} \\
    B_{\vec{k}}^\dagger & A^*(-\vec{k}) -C 
    \end{pmatrix} .
\end{align}
The $2 \times 2$ matrices $A(\vec{k})$, $B_{\vec{k}}$, and $C$ are defined in Ref.~\cite{Toth1} with the exchange couplings given above. Diagonalizing the Hamiltonian~\eqref{eq:Hspin_2} using a Bogoliubov-like transformation yields the magnon eigenmodes $\hat{\alpha}_{\vec{k}\nu}$, $\nu = 1, 2$,  with energies $\omega_{\nu}(\vec{k})$, respectively. The transformation is given by
\begin{align}
    \label{eq:Bogoliubov}
    \hat{b}_{\vec{k}j} = \sum_{\nu=1, 2} \left[Q_{j\nu}(\vec{k}) \hat{\alpha}_{\vec{k}\nu} + \bar{Q}_{j\nu}(\vec{k}) \hat{\alpha}^\dagger_{-\vec{k}\nu}\right] ;  
\end{align}
the coefficients $Q_{j\nu}(\vec{k})$ and $\bar{Q}_{j\nu}(\vec{k})$ are obtained from the decomposition of Eq.~(33) in Ref.~\cite{Toth1}. Substituting $\hat{s}_{\vec{R}} = 1 / \sqrt{N} \sum_{\vec{k}} e^{i\vec{k}\cdot\vec{R}} \hat{b}_{\vec{k}j}$ (for site $j = A, B$) and using the Bogoliubov transformation~\eqref{eq:Bogoliubov}, the effective Hamiltonian~\eqref{eq:Heff} can be expressed in terms of the magnon operators $\hat{\alpha}_{\vec{k}\nu}$, doublon operators $\hat{d}_{\vec{k}}$, and holon operators $\hat{e}_{\vec{k}}$. One finds
\begin{align}
    \label{eq:Heff_magnon}
    \hat{H}_\mathrm{eff} = -\frac{B_0}{\sqrt{N}} \sum_{\vec{k}, \vec{q}} \sum_{\nu=1,2} \left[\left(\hat{d}^\dagger_{\vec{k}A} \hat{d}_{\vec{k}-\vec{q}B} - \hat{e}^\dagger_{\vec{k}A} \hat{e}_{\vec{k}-\vec{q}B} \right) \left(Q_{A\nu}(\vec{q}) \hat{\alpha}_{\vec{q}\nu} + \bar{Q}_{B\nu}(\vec{q}) \hat{\alpha}^\dagger_{-\vec{q}\nu}\right) + (A\leftrightarrow B) \right] + \hat{H}_\mathrm{spin} \ .
\end{align}
The Hamiltonian~\eqref{eq:Heff_magnon} constitutes an interacting fermion-boson model, which in its full complexity is difficult to solve. To make further progress, we treat the doublons and holons as classical particles within a mean-field approximation. This is justified since in the experiments the photodoped carriers are generated incoherently by the pump pulse. Within this approximation, we replace the doublon and holon operators in Eq.~\eqref{eq:Heff_magnon} by their expectation values $\langle \hat{d}^\dagger_{\vec{k}A} \hat{d}_{\vec{k}^\prime B} \rangle = \sqrt{N} n_d \delta_{\vec{k}, \vec{k}^\prime}$ and $\langle \hat{e}^\dagger_{\vec{k}A} \hat{e}_{\vec{k}^\prime B} \rangle = \sqrt{N} n_h \delta_{\vec{k}, \vec{k}^\prime}$, where $n_d$ and $n_h$ are the doublon and holon densities, respectively. Since the exchange couplings have been chosen to reproduce the experimental magnon dispersion, we can neglect renormalization effects in the ground state. Thus, we only consider terms proportional to the time-dependent change of the doublon and holon densities. We then obtain the time-dependent magnon Hamiltonian
\begin{align}
    \label{eq:H_magnon_time}
    \hat{H}_\mathrm{mag}(t) = \sum_{\vec{q}, \nu} \omega_{\nu}(\vec{q}) \hat{\alpha}^\dagger_{\vec{q}\nu} \hat{\alpha}_{\vec{q}\nu} - B_0 \sum_{\vec{q}, \nu} \Delta n(t) \left[ \left(Q_{A\nu}(\vec{q}) + Q_{B\nu}(\vec{q}) \right) \hat{\alpha}_{\vec{q}\nu} + \left(\bar{Q}_{A\nu}(\vec{q}) + \bar{Q}_{B\nu}(\vec{q}) \right) \hat{\alpha}^\dagger_{-\vec{q}\nu} \right] \ .
\end{align}
Here, we introduced the total photodoped carrier density $\Delta n(t) = \Delta n_d(t)$, where $\Delta n_d(t)$ is the time-dependent changes in the doublon densities (the number of holons remains approximately constant).

Equation~\eqref{eq:H_magnon_time} describes how magnons are excited by the time-dependent photodoped carrier density $\Delta n(t)$. The solution of Eq.~\eqref{eq:H_magnon_time} is given in terms of a driven harmonic oscillator. Assuming that initially the magnons are in thermal equilibrium at temperature $T$, the time-dependent magnon occupation is given by 
\begin{align}
    f_\nu(\mathbf{q}, t) = f^\mathrm{eq}_\nu(\mathbf{q}) + |y_{\mathbf{q}\nu}(t)|^2 \ , 
\end{align}
where $f^\mathrm{eq}_\nu(\mathbf{q})$ is given by the equilibrium Bose-Einstein distribution and
\begin{align}
    \dot{y}_{\mathbf{q}\nu}(t) = - i \omega_\nu(\mathbf{q}) y_{\mathbf{q}\nu}(t) - i B_0 \Delta n(t) \left(\bar{Q}_{A\nu}(\vec{q}) + \bar{Q}_{B\nu}(\vec{q}) \right) \ .
\end{align}
For simplicity, the time-dependence of $\Delta n(t)$ is modeled as a smooth step function with a rise time of 25 fs, which is consistent with the simulations of the electronic dynamics described above. To estimate the effective strength of the spin-charge coupling, we compute the sublattice magnetization $S_{A,B}$ per Cu site after photoexcitation using Eq. (50) in Ref.~\cite{Toth1}. For the spin-1/2 model considered here, $S^\mathrm{eq}_A = 0.39 \hbar$ in equilibrium at $T = 190$ K. We then chose the amplitude of $B_0 \Delta n(t)$ to reproduce the experimentally observed reduction of the AFM order by approximately 50\%.

\begin{figure}[ht]
    \centering
    \includegraphics[width=0.9\textwidth]{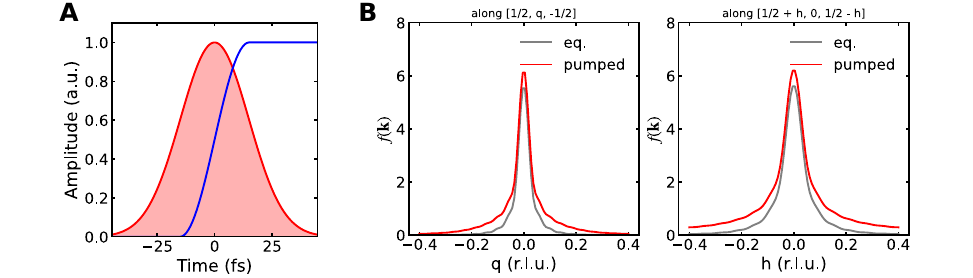}
    \caption{\textbf{Photoinduced magnon occupation.} (\textbf{A}) Sketch of the envelope of the pump pulse and the approximated $B_0 \Delta n(t)$. (\textbf{B}) Magnon occupation $f(\vec{k})$ along the reciprocal axes in equilibrium at $T=190$ K, and immediately after photoexcitation.
    The wiggles in the equilibrium distribution are due to Fourier interpolation artifacts.}
    \label{fig:magnon_occupation}
\end{figure}

The resulting magnon occupation after photoexcitation (Fig.~\ref{fig:magnon_occupation}\textbf{B}) is significantly non-thermal, with a broad distribution of magnons excited across the entire Brillouin zone. Similar non-thermal magnon distributions have been reported in Ref.~\cite{Mazzone1}.

\subsubsection*{\underline{Theory of magnon-magnon scattering}}

After their initial excitation, the magnons undergo magnon-magnon scattering processes, redistributing energy and momentum until a thermal magnon distribution is reached. To model this thermalization process, we employ a quantum Boltzmann equation (QBE) for the magnon occupation $f(\vec{k}, t)$:
\begin{align}
    \label{eq:qbe}
    \frac{d}{dt} f_\nu(\vec{k}, t) = S_\nu[f](\vec{k}, t) \ .
\end{align}
Note that the two magnon branches $\nu = 1, 2$ are degenerate in our simplified model.
The magnon-magnon scattering term $S_\nu[f](\vec{k}, t)$ is derived by going to fourth order in the Holstein-Primakoff expansion of the underlying Heisenberg model, as demonstrated in Ref.~\cite{Kalthoff1} for the square lattice. Here we generalize this approach to the present case of CuO with the exchange couplings introduced above. To this end, we start from Heisenberg Hamiltonian (now $\vec{R}$ runs over all lattice vectors in the magnetic unit cell, while $j, j^\prime = A, B$ denote the two sublattices):
\begin{align}
    \hat{H}_\mathrm{Heis} = \sum_{\langle \vec{R}, \vec{R}^\prime \rangle} \sum_{j j^\prime} J_{j j^\prime} (\vec{R} - \vec{R}^\prime) \hat{\vec{S}}_{\vec{R} j} \cdot \hat{\vec{S}}_{\vec{R}^\prime j^\prime} =  \sum_{\langle \vec{R}, \vec{R}^\prime \rangle} \sum_{j j^\prime}  J_{j j^\prime}(\vec{R} - \vec{R}^\prime) \hat{\vec{S}}^\prime_{\vec{R}i} \vec{V}^T_{j} \vec{V}_{j^\prime} \hat{\vec{S}}^\prime_{\vec{R}^\prime j^\prime} \ ,
\end{align}
where $\hat{\vec{S}}^\prime_{\vec{R}} = \vec{V}^T_{j} \hat{\vec{S}}_{\vec{R}}$ are the spin operators in the rotated frame where all spins are aligned.
It is convenient to express the rotation matrices $\vec{V}_{j}$ in terms the vectors $\vec{u}_j$, $\vec{v}_j$ as in Ref.~\cite{Toth1}. Expanding the Holstein-Primakoff transformation up to fourth order and defining $J^\prime_{j, j^\prime}(\vec{R}) = \vec{V}^T_{j}  J_{j j^\prime}(\vec{R}) \vec{V}_{j^\prime}$, we obtain the quartic magnon Hamiltonian 
\begin{align}
    \label{eq:mag_mag_interaction}
    \hat{H}_\mathrm{mag-mag} = - \frac{1}{8} \sum_{\vec{R}, \vec{R}^\prime} \sum_{j, j^\prime} J^\prime_{j j^\prime}(\vec{R} - \vec{R}^\prime) \Bigg[ &\vec{u}^\dagger_j \vec{u}^*_{j^\prime} \hat{b}_{\vec{R}j} \hat{b}^\dagger_{\vec{R}^\prime j^\prime} \hat{b}_{\vec{R}^\prime j^\prime} \hat{b}_{\vec{R}^\prime j^\prime} + \vec{u}^\dagger_{j} \vec{u}_{j^\prime} \hat{b}_{\vec{R}j} \hat{b}^\dagger_{\vec{R}^\prime j^\prime} \hat{b}^\dagger_{\vec{R}^\prime j^\prime} \hat{b}_{\vec{R}^\prime j^\prime}  \nonumber  \\
    &+  \vec{u}^\dagger_j \vec{u}^*_{j^\prime}  \hat{b}^\dagger_{\vec{R}j}  \hat{b}_{\vec{R}j} \hat{b}_{\vec{R}j} \hat{b}_{\vec{R}^\prime j^\prime} 
    + \vec{u}^\dagger_{j} \vec{u}_{j^\prime}  \hat{b}^\dagger_{\vec{R}j}  \hat{b}_{\vec{R}j} \hat{b}_{\vec{R}j}  \hat{b}^\dagger_{\vec{R}^\prime j^\prime} + \mathrm{H.c.} \Bigg] \ .
\end{align}
Fourier transforming Eq.~\eqref{eq:mag_mag_interaction} yields $\hat{H}_\mathrm{mag-mag} = - (\hat{V}^{(1)} + \hat{V}^{(2)} + \mathrm{H.c.})/4$ with
\begin{align}
    \label{eq:mag_mag_interaction_k}
    \hat{V}^{(1)} = \frac{1}{N} \sum_{j, j^\prime} \sum_{\vec{k}_1, \vec{k}_2, \vec{k}_3, \vec{k}_4} \delta_{\vec{k}_1, \vec{k}_2 - \vec{k}_3 -\vec{k}_4} A^*_{j j^\prime}(\vec{k}_1) \hat{b}_{\vec{k}_1 j} \hat{b}^\dagger_{\vec{k}_2 j^\prime} \hat{b}_{\vec{k}_3 j^\prime} \hat{b}_{\vec{k}_4 j^\prime} \ , \\
    \hat{V}^{(2)} = \frac{1}{N} \sum_{j, j^\prime} \sum_{\vec{k}_1, \vec{k}_2, \vec{k}_3, \vec{k}_4} \delta_{\vec{k}_1, \vec{k}_2 + \vec{k}_3 -\vec{k}_4} B_{j j^\prime}(\vec{k}_1) \hat{b}_{\vec{k}_1 j} \hat{b}^\dagger_{\vec{k}_2 j^\prime} \hat{b}^\dagger_{\vec{k}_3 j^\prime} \hat{b}_{\vec{k}_4 j^\prime} \ ,
\end{align}
where we have defined the coefficients $A_{j j^\prime}(\vec{k}) = \vec{u}^T_j \vec{u}_{j^\prime} J_{jj^\prime}(-\vec{k})$ and $B_{j j^\prime}(\vec{k}) = \vec{u}^\dagger_j \vec{u}_{j^\prime} J_{jj^\prime}(\vec{k})$. Finally, substituting the transformation~\eqref{eq:Bogoliubov} into Eqs.~\eqref{eq:mag_mag_interaction_k}, we obtain the magnon-magnon interaction in terms of the magnon eigenmodes $\hat{\alpha}_{\vec{k}\nu}$: 
\begin{align}
    \label{eq:mag_mag_interaction_final}
    \hat{H}_\mathrm{mag-mag} = \frac{1}{N^2}\sum_{\vec{k}_1, \vec{k}_2, \vec{k}_3, \vec{k}_4} \sum_{\nu_1, \nu_2, \nu_3, \nu_4} V_{\nu_1 \nu_2 \nu_3 \nu_4}(\vec{k}_1, \vec{k}_2, \vec{k}_3, \vec{k}_4) \hat{\alpha}^\dagger_{\vec{k}_1 \nu_1} \hat{\alpha}^\dagger_{\vec{k}_2 \nu_2} \hat{\alpha}_{\vec{k}_3 \nu_3} \hat{\alpha}_{\vec{k}_4 \nu_4} \delta_{\vec{k}_1 + \vec{k}_3, \vec{k}_2 + \vec{k}_4} \ .
\end{align}
Using the magnon-magnon vertex $V_{\nu_1 \nu_2 \nu_3 \nu_4}(\vec{k}_1, \vec{k}_2, \vec{k}_3, \vec{k}_4)$ from Eq.~\eqref{eq:mag_mag_interaction_final} for the degenerate mode $\nu_1=\nu_2=\nu_3=\nu_4$, the magnon-magnon scattering term in the QBE is given by
\begin{align}
    S_\nu[f](\vec{k}_1, t) = \frac{2\pi}{N^2} \sum_{\vec{k}_2, \vec{k}_3, \vec{k}_4} & \delta_{\vec{k}_1 + \vec{k}_2, \vec{k}_3 + \vec{k}_4}|V_{\nu \nu \nu \nu}(\vec{k}_1, \vec{k}_2, \vec{k}_3, \vec{k}_4)|^2 \delta(\omega_{\nu}(\vec{k}_1) + \omega_{\nu}(\vec{k}_2) - \omega_{\nu}(\vec{k}_3) - \omega_{\nu}(\vec{k}_4)) \nonumber \\ & \times \left[\bar{f}_\nu(\vec{k}_1, t) \bar{f}_\nu(\vec{k}_2, t) f_\nu(\vec{k}_3, t) f_\nu(\vec{k}_4, t) 
    - f_\nu(\vec{k}_1, t) f_\nu(\vec{k}_2, t) \bar{f}_\nu(\vec{k}_3, t) \bar{f}_\nu(\vec{k}_4, t) \right] \ .
\end{align}
Here, $\bar{f}_\nu(\vec{k}, t) = 1 + f_\nu(\vec{k}, t)$. We solve the QBE~\eqref{eq:qbe} numerically using a discretization of the Brillouin zone with $N = 32 \times 48 \times 32$ $\vec{k}$-points. The Dirac-delta function is approximated by a biweight kernel with a width of 3.0\, meV. The time-stepping of Eq.~\eqref{eq:qbe} is performed using an adaptive Dormand-Prince Runge-Kutta solver.

\begin{figure}[ht]
    \centering
    \includegraphics[width=0.8\textwidth]{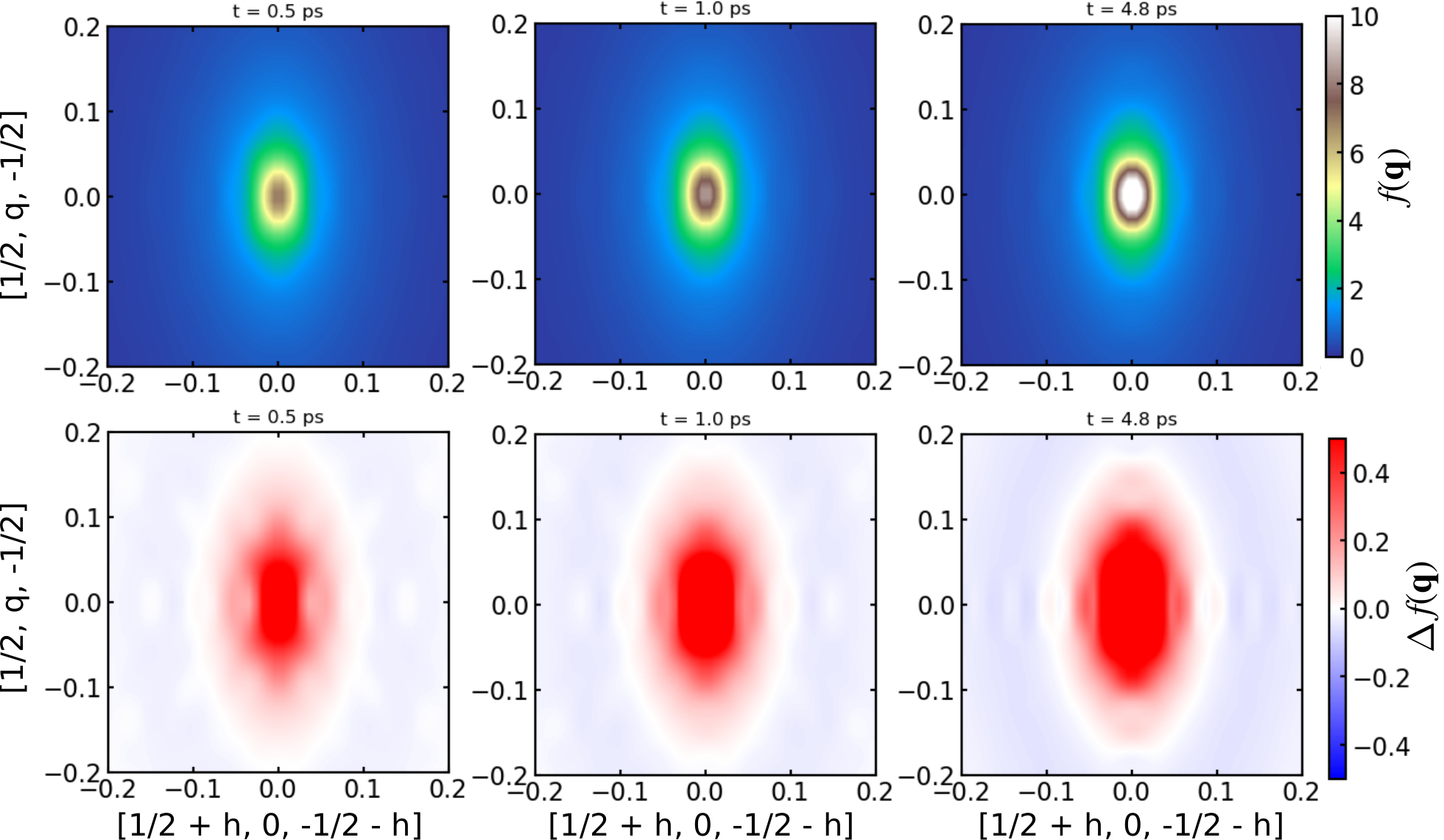}
    \caption{\textbf{Magnon thermalization dynamics.} Snapshots of the magnon distribution $f(\mathbf{q},t)$ (top row) and difference $\Delta f(\vec{q}, t) = f(\mathbf{q}, t) - f(\mathbf{q}, 0)$ (bottom row) along the reciprocal-space directions equivalent to the experiments for different time steps. \label{fig:magnon_evol}}
\end{figure}

Figure~\ref{fig:magnon_evol} shows the time-dependent distribution for a few representative time steps. Initially, the distribution shows broad tails extending to high energies. As time progresses, magnon-magnon scattering leads to a redistribution of magnons towards lower energies. Due the large phase-space volume for magnon-magnon scattering in the three-dimensional system, the magnon distribution close to $\Gamma$ grows quickly. 

\begin{figure}[ht]
    \centering
    \includegraphics[width=0.8\textwidth]{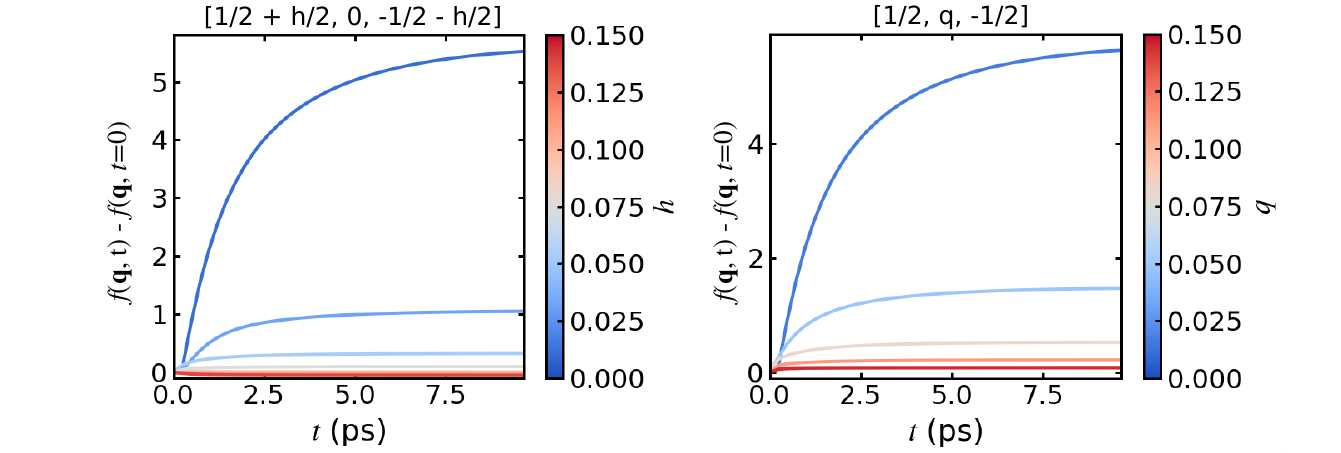}
    \caption{\textbf{Momentum-dependent magnon occupation dynamics.} Time evolution of the change of magnon occupation $\Delta f(\vec{q}, t) = f(\vec{q}, t) - f(\vec{q}, 0)$ for selected $\vec{q}$-points, indidcated by the color coding. 
    \label{fig:time_evol_cuts}}
\end{figure}

To investigate the momentum dependence of the thermalization dynamics in more detail, we plot the time-dependent change of the magnon occupation $\Delta f(\vec{q}, t)$ for selected $\vec{q}$-points in Fig.~\ref{fig:time_evol_cuts}, similar to the experimental analysis. The results show that the magnon occupation shows that away from the magnetic $\Gamma$-point, the population dynamics saturates quickly. Closer to $\Gamma$, however, the magnon occupation continues to grow even at later times, consistent with the experimental observations.

\subsubsection*{\underline{Theory of magnon-phonon scattering}}

We have also investigated the role of magnon-phonon scattering for the recovery dynamics. To this end, we analyzed the scattering term $S^\mathrm{mp}_\nu[f](\vec{k}, t)$ due to magnon-phonon interactions within the QBE~\eqref{eq:qbe}. To derive a microscopic model of magnon-phonon coupling, we start from the Heisenberg Hamiltonian with exchange couplings $J_{jj^\prime}(\vec{R} - \vec{R}^\prime)$ that depend on the atomic positions. Assmining the magnetic exchange couplings depend on the relative distance between two Cu sites, we expand $J_{jj^\prime}(\vec{R} - \vec{R}^\prime)$ to linear order in the atomic displacements $\vec{u}_{\vec{R} j}$:
\begin{align}
    J_{jj^\prime}(\vec{R} - \vec{R}^\prime) \approx J_{jj^\prime}^0(\vec{R} - \vec{R}^\prime) + \vec{\nabla} J_{jj^\prime}^0(\vec{R} - \vec{R}^\prime) \cdot \left(\vec{u}_{\vec{R} j} - \vec{u}_{\vec{R}^\prime j^\prime} \right) \ ,
\end{align}
where $J_{jj^\prime}^0(\vec{R} - \vec{R}^{ \prime})$ are the equilibrium exchange couplings. This yields the magnon-phonon interaction Hamiltonian
\begin{align},
    \label{eq:mag_phonon_interaction}
    \hat{H}_\mathrm{mag-ph} = \sum_{\langle \vec{R}, \vec{R}^\prime \rangle} \sum_{j j^\prime} \vec{\nabla} J_{jj^\prime}^0(\vec{R} - \vec{R}^\prime) \cdot \left(\vec{u}_{\vec{R} j} - \vec{u}_{\vec{R}^\prime j^\prime} \right) \hat{\vec{S}}_{\vec{R} j} \cdot \hat{\vec{S}}_{\vec{R}^\prime j^\prime} \ .
\end{align}
The atomic displacements can be expressed in terms of phonon modes as 
\begin{align}
    \vec{u}_{\vec{R} j} = \frac{1}{\sqrt{N}} \sum_{\vec{k}, \lambda} \frac{1}{\sqrt{2 M_j \Omega_\lambda(\vec{k})}} \vec{e}_{j \lambda}(\vec{k}) \left( \hat{a}_{\vec{q} \lambda} e^{i \vec{k} \cdot (\vec{R} + \vec{r}_j)} + \hat{a}^\dagger_{-\vec{k} \lambda} e^{-i \vec{k} \cdot (\vec{R} + \vec{r}_j)} \right) \ ,
\end{align}
where $\hat{a}_{\vec{k} \lambda}$ ($\hat{a}^\dagger_{\vec{k} \lambda}$) are phonon annihilation (creation) operators for mode $\lambda$ with momentum $\vec{k}$, frequency $\Omega_\lambda(\vec{k})$, polarization vector $\vec{e}_{j \lambda}(\vec{k})$, and mass $M_j$ of atom $j$ in the unit cell. Next, we express the spin operators in Eq.~\eqref{eq:mag_phonon_interaction} in terms of magnon operators using the Holstein-Primakoff transformation and the Bogoliubov transformation~\eqref{eq:Bogoliubov}. Collecting all terms that are second order in magnon operators and first order in phonon operators, we obtain two distinct magnon-phonon interaction processes: (i) normal scattering (magnon number-conserving), where a magnon scatters from momentum $\vec{q}$ to $\vec{q} + \vec{k}$ by absorbing or emitting a phonon with momentum $\vec{k}$, and (ii) anomalous scattering (magnon pair creation/annihilation), where two magnons with momenta $\vec{q}$ and $-\vec{q} + \vec{k}$ are created or annihilated by absorbing or emitting a phonon with momentum $\vec{k}$. The respective interaction Hamiltonians read as
\begin{align}
    \hat{H}_\mathrm{mag-ph}^\mathrm{normal} &= \frac{1}{\sqrt{N}} \sum_{\vec{q}, \vec{k}} \sum_{\nu} \sum_{\lambda} 
    \frac{g^\mathrm{normal}_{\nu\lambda}(\vec{q}, \vec{k})}{\sqrt{2 M \Omega_\lambda(\vec{k})}} \left( \hat{a}_{\vec{k} \lambda} + \hat{a}^\dagger_{-\vec{k} \lambda} \right) \hat{\alpha}^\dagger_{\vec{q} + \vec{k} \nu} \hat{\alpha}_{\vec{q} \nu} + \mathrm{h.c.} \ ,
\end{align}
\begin{align}
    \hat{H}_\mathrm{mag-ph}^\mathrm{anomalous} &= \frac{1}{\sqrt{N}} \sum_{\vec{q}, \vec{k}} \sum_{\nu} \sum_{\lambda} 
    \frac{g^\mathrm{anomalous}_{\nu\lambda}(\vec{q}, \vec{k})}{\sqrt{2 M \Omega_\lambda(\vec{k})}}  \hat{a}^\dagger_{\vec{k} \lambda} \hat{\alpha}_{\vec{q} + \vec{k} \nu} \hat{\alpha}_{-\vec{q} \nu} +  \mathrm{h.c.} \ .
\end{align}
The calculation of the magnon-phonon coupling constants $g^\mathrm{normal}_{\nu\lambda}(\vec{q}, \vec{k})$ and $g^\mathrm{anomalous}_{\nu\lambda}(\vec{q}, \vec{k})$ involves lengthy expressions depending on the exchange gradients $\vec{\nabla} J_{jj^\prime}^0(\vec{R} - \vec{R}^\prime)$, the Bogoliubov transformation coefficients, and the phonon polarization vectors. We concentrate here on the phase space structure of magnon-phonon scattering and simplify the coupling constants by (i) replacing $\vec{e}_{j\lambda}(\vec{k}) \cdot \vec{\nabla} J_{jj^\prime}^0(\vec{R} - \vec{R}^\prime) \rightarrow c^x$, where $c^x$ ($x=$normal, anomalous) are constants, and (ii) approximating the effective masses $M_j$ by a single average mass $M$ of copper.

Using Fermi's golden rule, we obtain the magnon-phonon scattering term in the QBE~\eqref{eq:qbe} as
\begin{align}
    \label{eq:normal_magnon_phonon}
    S^\mathrm{mp, normal}_\nu[f](\vec{q}) = \frac{2\pi}{N} \sum_{\vec{k}, \lambda} & \frac{|g^\mathrm{normal}_{\nu\lambda}(\vec{q}, \vec{k})|^2}{2 M \Omega_\lambda(\vec{k})} \Bigg\{ \delta(\omega_\nu(\vec{q} + \vec{k}) - \omega_\nu(\vec{q}) - \Omega_\lambda(\vec{k})) \nonumber \\ & \times \left[ \bar{f}_\nu(\vec{q} + \vec{k}) f_\nu(\vec{q}) N_\lambda(\vec{k}) - f_\nu(\vec{q} + \vec{k}, t) \bar{f}_\nu(\vec{q}, t) \bar{N}_\lambda(\vec{k}) \right] \nonumber \\
    + & \delta(\omega_\nu(\vec{q} + \vec{k}) - \omega_\nu(\vec{q}) + \Omega_\lambda(\vec{k})) \nonumber \\ & \times \left[ \bar{f}_\nu(\vec{q} - \vec{k}, t) f_\nu(\vec{q}, t) \bar{N}_\lambda(\vec{k}) - f_\nu(\vec{q} - \vec{k}, t) \bar{f}_\nu(\vec{q}, t) N_\lambda(\vec{k}) \right] \Bigg\} \ ,
\end{align}
where $N_\lambda(\vec{q})$ is the phonon occupation number for mode $\lambda$ with momentum $\vec{q}$, and where $\bar{N}_\lambda(\vec{q}) = 1 + N_\lambda(\vec{q})$. The anomalous magnon-phonon scattering term reads as
\begin{align}
    \label{eq:anomalous_magnon_phonon}
    S^\mathrm{mp, anomalous}_\nu[f](\vec{q}) = \frac{2\pi}{N} \sum_{\vec{k}, \lambda} & \frac{|g^\mathrm{anomalous}_{\nu\lambda}(\vec{q}, \vec{k})|^2}{2 M \Omega_\lambda(\vec{k})} \delta(\omega_\nu(\vec{q} + \vec{k}) + \omega_\nu(-\vec{q}) - \Omega_\lambda(\vec{k})) \nonumber \\ & \times \left[ \bar{f}_\nu(\vec{q} + \vec{k}, t) \bar{f}_\nu(-\vec{q}, t) N_\lambda(\vec{k}) - f_\nu(\vec{q} + \vec{k}, t) f_\nu(-\vec{q}, t) \bar{N}_\lambda(\vec{k}) \right] \ .
\end{align}

The dispersion of the acoustic phonons $\Omega_\lambda(\vec{k})$ are estimated from the model 
\begin{align}
    \Omega_\lambda(\vec{k}) = \left(\sum^3_{i=1} C^{(0)}_{i\lambda} \sin^2\left(\frac{\vec{k}\cdot \vec{a}_i}{2}\right) \right)^{1/2} \ .
\end{align}
Here, $\vec{a}_i$ are the primitive lattice vectors of CuO, and $C^{(0)}_{i\lambda}$ are fitting parameters chosen to reproduce the acoustic phonon velocities along high-symmetry directions from Ref.~\cite{Reichardt1}. Similarly, we approximate the optical phonon modes by
\begin{align}
    \Omega_\lambda(\vec{k}) = C^{(1)}_\lambda + \sum^3_{i=1} C^{(2)}_{i\lambda} \cos(\vec{k} \cdot \vec{a}_i) \ ,
\end{align}
where $C^{(1)}_\lambda$ and $C^{(2)}_{i\lambda}$ are fitting parameters chosen to reproduce the optical phonon frequencies reported in ref.~\cite{Reichardt1} at the $\Gamma$-point and at the zone boundaries.

With these ingredients, we evaluated Eqs.~\eqref{eq:normal_magnon_phonon} and~\eqref{eq:anomalous_magnon_phonon} numerically on a $N = 64\times 64 \times 64$ $\vec{k}$-point grid. We fixed $c^x = 1$ eV/\AA\ for both normal and anomalous magnon-phonon coupling constants. The phonon occupation numbers $N_\lambda(\vec{k})$ are defined by the Bose-Einstein distribution at $T=200$\,K (corresponding to a slightly excited distribution with respect to the equilibrium), while we used the steady-state distribution from the magnon-magnon QBE simulation as initial magnon distribution. For simplicity, we only considered the initial scattering rate $S^\mathrm{mp}_\nu[f](\vec{q}, t=0)$ due to magnon-phonon interactions, from which we can estimate the magnon-phonon scattering rate as $\Gamma_\nu(\vec{q}) = S^\mathrm{mp}_\nu[f](\vec{q}, t=0)/f_\nu(\vec{q}, t=0)$. 

\begin{figure}[ht]
    \centering
    \includegraphics[width=\textwidth]{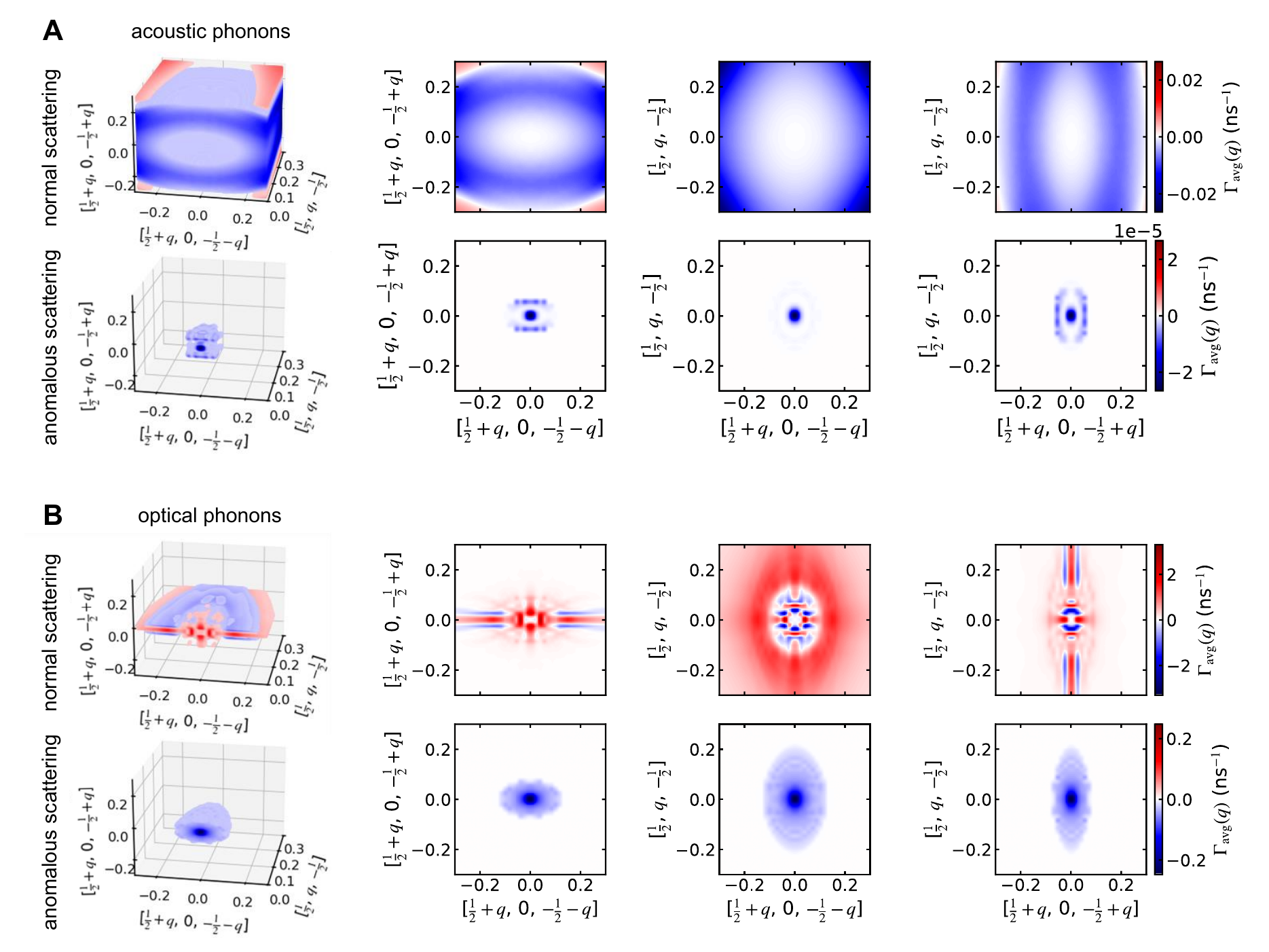}
    \caption{\textbf{Magnon-phonon scattering rate.} Mode-averaged magnon-phonon scattering rate for (\textbf{A}) the acoustic phonons and (\textbf{B}) the lowest-energy optical phonons in three-dimensional momentum space and along high-symmetry planes. Top row: normal scattering, bottom row: anomalous scattering. 
    \label{fig:magnon_phonon_scattering}}
\end{figure}

The results for the acoustic phonons and the lowest-energy optical phonons are shown in Fig.~\ref{fig:magnon_phonon_scattering}. While normal magnon-phonon scattering is an order of magnitude faster than anomalous scattering, the normal process cannot reduce the total magnon population. There are redistribution processes in momentum space due to the different anisotropy of the magnon and the phonon dispersion, mostly at momenta away from the magnetic $\Gamma$-point. Close to $\Gamma$, however, the normal magnon-phonon scattering rate is strongly suppressed due to the lack of phase space for energy- and momentum conservation. The anomalous magnon-phonon scattering rate is purely negative and is generally smaller in magnitude. It is most effective close to the magnetic $\Gamma$-point, where two low-energy magnons can be annihilated by emitting a phonon. We conclude that only the anomalous magnon-phonon scattering process can contribute to the magnetic recovery dynamics observed in the experiments. The small magnitude of the anomalous magnon-phonon scattering rate, originating from the mismatch of magnon and phonon dispersion, is consistent with the experimentally observed slow recovery dynamics. 

We also evaluated the magnon-phonon scattering rates for acoustic phonons. The corresponding scattering rates are orders of magnitude smaller than those for the lowest-energy optical phonons, due to the small energy scale of acoustic phonons compared to magnons. 

\subsubsection*{\underline{Theory of RDS of magnon origin}}

The magnetic diffuse scattering intensity is calculated based on Kramers-Heisenberg equation~\cite{Ament1}. At $T=0$, only the ground state is taken into account as initial state. Sacttering intensity with incoming (outgoing) photon momentum, photon energy, and polarization $\mathbf{k}_i,\omega_i,\boldsymbol{\epsilon}_i$ ($\mathbf{k}_o,\omega_o,\boldsymbol{\epsilon}_o$) can be written as
 \begin{align}
    \label{eq:kramers-heisenberg2}
    I_{\mathrm{DS}} (\mathbf{k}_i,\mathbf{k}_o,\omega_i,\omega_o,\epsilon_i,\epsilon_o) \nonumber = \sum_f |\langle f| \hat{D}_o^{\dagger} \frac{1}{\omega_i + E_g +i\tau_n - \hat{H}_n} \hat{D}_i | g\rangle|^2 \nonumber  L_{\tau_f} (\omega_i - \omega_o + E_g - E_f) \ , \\
\end{align}
where $\hat{D}_i$ ($\hat{D}_o$) is light-matter dipole operator for incoming (outgoing) photon, which excite one core electron into the valence orbital. $|g\rangle$ is the initial state (before scattering) with energy $E_g$ and $|f\rangle$ is the final state (after scattering) with energy $E_f$. The propagator in between dipole operators uses Hamiltonian $\hat{H}_n$ in the presence of one excited core-hole and a corresponding lifetime $\tau_n$. $L_{\tau_f}(E)$ is a Lorentzian with lifetime broadening $\tau_f$. We simplify the scattering process by defining a local transition operator $\hat{T}_{io}^{\mathbf{R}}$~\cite{Haverkort1} as
    \begin{align}
        \hat{T}_{io}^{\mathbf{R}} &\equiv \hat{D}_o^{\mathbf{R}\dagger} \frac{1}{\omega_i + E_g +i\tau_n - \hat{H}_n} \hat{D}^{\mathbf{R}}_i \ .
    \end{align}
Here the local dipole operator $\hat{D}^{\mathbf{R}}$ ($\hat{D}^{\mathbf{R}\dagger}$) annihilate (create) a core-hole at site $\mathbf{R}$. The local transition operator $\hat{T}^\mathbf{R}_{io}$ is defined so that the incoming and outgoing photons act on the same site $\mathbf{R}$, and its Fourier transformation is defined as $\hat{T}_{io}^{\mathbf{q}} = \sum_{\mathbf{R}} e^{i \mathbf{q} \cdot  \mathbf{R}} \hat{T}_{io}^{\mathbf{R}}$. We now restrict the possible excitations ($|l\rangle$) to be within a small energy range $\omega_i-\omega_o \approx 0$ on the scale of the electronic spectrum. The cross-section is then separated into an \textit{elastic} and \textit{inelastic} part as
\begin{align}
    \label{eq:RIXS_nearly_resonant}
        I_{\mathrm{DS}}(\mathbf{q}=\mathbf{k}_i-\mathbf{k}_o ,&\omega_i-\omega_o \rightarrow 0) 
        \approx  |\langle g | \hat{T}^{\mathbf{q}}_{io} | g\rangle|^2 \nonumber \\+ &\sum_{l \neq g,E_l-E_g \rightarrow0} \langle g| \hat{T}_{io}^{\mathbf{q}\dagger} | l\rangle \langle l | \hat{T}^{\mathbf{q}}_{io} | g\rangle L_{\tau_f}(\omega_i - \omega_o + E_g - E_l) ,
\end{align}
where in the second line summation runs over low energy excitation $|l\rangle$ close to the ground states. The form of transition operator $\hat{T}^{\mathbf{R}}_{io}$ can be related to the optical conductivity tensor~\cite{Haverkort1,Newton1}. If site $\mathbf{R}$ has local magnetization~\cite{Haverkort1}, the matrix elements of $\hat{T}_{io}^{\mathbf{q}}$ in Eq.~\eqref{eq:RIXS_nearly_resonant} can be approximated as matrix elements of a series of quantum spin operators~\cite{Haverkort2,Haverkort1}.  Here, we approximate the matrix element expression to the first order, i.e. one spin operator for one $\hat{T}_{io}^\mathbf{R}$. In our case of magnetic diffuse scattering in CuO, we approximate the point group symmetries of all Cu sites to be spherical symmetry for simplification. The time-dependent scattering signal is then calculated by the time-dependent magnon distribution, simulated with QBE. Now we denote the time-dependent excited magnon distribution as $|g(t)\rangle$. For incoming (outgoing) light polarization $\boldsymbol{\epsilon}_i$ ($\boldsymbol{\epsilon}_o$), the first elastic term in the right hand side of  Eq.~\eqref{eq:RIXS_nearly_resonant} can be simplified as
\begin{align}
    \label{eq:magnetic_bragg}
    I_{io}^{\mathrm{elastic}}(t) = |\langle g(t) | \hat{T}^{\mathbf{q}}_{io} | g(t)\rangle|^2 \sim | \langle g(t)|  \sum_{\alpha=x,y,z} c_{\alpha} \sum_{\mathbf{R}, j} e^{i\mathbf{q} \cdot  (\mathbf{R} + \mathbf{r}_j)} \hat{S}^{\alpha}_{\mathbf{R}j} |g(t) \rangle|^2 = |\langle g(t)| \sum_{\alpha} c_{\alpha} \hat{S}^{\alpha}(\mathbf{q}) |g(t) \rangle|^2 \ ,
\end{align}
where the coefficient $c_{\alpha}=[\boldsymbol{\epsilon}_i \times \boldsymbol{\epsilon}_o]_{\alpha}$ reflects the experimental scattering geometry.  The Fourier transformation of spin operator is defined as $\hat{S}^{\alpha}(\mathbf{q})=\sum_{\mathbf{R}}e^{i\mathbf{q}\cdot \mathbf{R}}\hat{S}_{\mathbf{R}}^{\alpha}$. Equation~\eqref{eq:magnetic_bragg} measures the \textit{magnetic Bragg peak} with extra coefficient $c_{\alpha}$. Similarly, the second inelastic term in Eq.~\eqref{eq:RIXS_nearly_resonant} reads as
\begin{align}
    \label{eq:inelastic_correlators}
    I_{io}^{\mathrm{inelastic}}(t) \propto \sum_{l}\langle g(t)| \hat{T}_{io}^{\mathbf{q}\dagger} | l\rangle \langle l | \hat{T}^{\mathbf{q}}_{io} | g(t)\rangle 
    &\sim \sum_{\alpha \alpha'} c_{\alpha}^{*} c_{\alpha'}\sum_l \langle g(t)| \hat{S}^{\alpha}(\mathbf{-q}) |l \rangle  \langle l| \hat{S}^{\alpha'}(\mathbf{q}) |g(t) \rangle \nonumber \\
    &= \frac{1}{N} \sum_{\mathbf{R}\mathbf{R'}} \sum_{j j^\prime} e^{i \mathbf{q} \cdot (\mathbf{R} - \mathbf{R}' + \mathbf{r}_j + \mathbf{r}_{j'})} \sum_{\alpha \alpha'} F_{\alpha \alpha'} \langle g(t)| \hat{S}^{\alpha}_{\mathbf{R}j} \hat{S}^{\alpha'}_{\mathbf{R'} j'} |g(t) \rangle \ ,
\end{align}
where $N$ refers to the number of unit cells and where $j, j'$ run over the sublattice sites. The last term in Eq.~\eqref{eq:inelastic_correlators} defines the spin-spin correlation function
\begin{align}
    S_{\alpha \alpha^\prime}(\mathbf{q}, t) =  \frac{1}{N} \sum_{\mathbf{R}\mathbf{R'}} \sum_{j j^\prime} e^{i \mathbf{q} \cdot (\mathbf{R} - \mathbf{R}' + \mathbf{r}_j + \mathbf{r}_{j'})}
    \langle g(t) | \hat{S}^{\alpha}_{\mathbf{R}j} \hat{S}^{\alpha'}_{\mathbf{R'} j'} |g(t) \rangle \ .
\end{align}
Hence, the inelastic scattering signal~\eqref{eq:inelastic_correlators} directly reveals the spin-spin correlation function 

\begin{figure}[ht]
    \centering
    \includegraphics[width=0.9\textwidth]{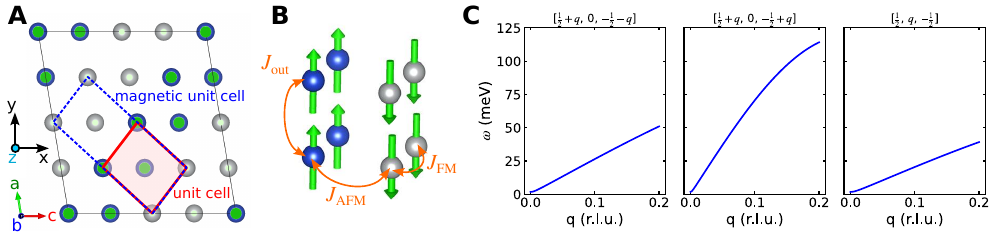}
    \caption{\textbf{Model and magnon dispersion.} (\textbf{A}) Simplified (magnetic) unit cell. The spin orientation is described in the $xyz$ coordinate system. (\textbf{B}) Model unit cell with next-nearest-neighbor exchange couplings. (\textbf{C}) Magnon dispersion of the model spin Hamiltonian. The dispersion was computed along three directions in the Brillouin zone corresponding to the simplified unit cell and convert to (approximately) to the reciprocal path directions used in the experiments.}
    \label{fig:spin_model}
\end{figure}

Equation~\eqref{eq:inelastic_correlators} is exactly spin structural factor with extra prefactor $F_{\alpha\alpha'}=c_{\alpha}^*c_{\alpha'}$, which is determined by the scattering geometry. Next, the spin operators in Eq.~\eqref{eq:inelastic_correlators} are expressed with bosonic operators with Holstein-Primakoff transformation, and further transformed to magnon eigenmodes as in Eq.~\eqref{eq:Bogoliubov}. 
We obtain
\begin{align}
    \label{eq:inelastic_magnon}
    I_{io}^{\mathrm{inelastic}}(t) \propto \sum_{\alpha \alpha'} F_{\alpha \alpha'} \sum_{\nu} \left[ A_\nu(\vec{q}) f_\nu(\vec{q}, t)
        + \bar{A}_\nu(\vec{q}) \bar{f}_\nu(\vec{q}, t) \right]  \ ,
\end{align}
where the coefficients $A_\nu(\vec{q})$,  and $\bar{A}_\nu(\vec{q})$ are calculated from the transformation coefficients $Q_{j\nu}$ and $\bar{Q}_{j\nu}$ in Eq.~\eqref{eq:Bogoliubov}, and where $\bar{f}_\nu = f_\nu + 1$. Equation~\eqref{eq:inelastic_magnon} can be directly evaluated based on time-dependent magnon distribution simulated from QBE. We include both $s$ and $p$ polarizations in the geometric coefficient $F_{\alpha\alpha'}=c_{\alpha}^{s*}c_{\alpha}^{s} + c_{\alpha}^{p*}c_{\alpha}^{p}$, and they are calculated based on experimental geometry as 
\begin{align}
    \label{eq:geoetric_factors}
    &F_{xx}= 0.505, F_{xy}=0.453, F_{xz}=-0.130 , \nonumber\\ 
    &F_{yx}=0.453, F_{yy}=0.415, F_{yz}=-0.037 , \nonumber\\ 
    &F_{zx}=-0.130, F_{xy}=-0.037, F_{zz}=0.822 . 
\end{align}
The coordinate system for the light polarization is the same as for the spin orientation, sketched in Fig.~\ref{fig:spin_model}\textbf{A}.

% The two main sections of the supplement can be split up using headings.
% If your supplement is very short you might need to uncomment the following line to avoid
% layout problems with the figures and tables.

\newpage

%%%%%%%%%%%%%%%% SUPPLEMENTARY FIGURES %%%%%%%%%%%%%%%

% \begin{figure} 
% 	\centering
% 	\includegraphics[width=\textwidth]{FigS9.pdf}
% 	\caption{\textbf{Comparison of resonant X-ray scattering profiles around ($\mathbf{\nicefrac{1}{2}}$ 0 $\mathbf{-\nicefrac{1}{2}}$) before and after photoexcitation (3 ps).}
%         The profiles were extracted from the tr-RDS data shown in Fig.~\ref{fig:3}\textbf{B}. Thus, the diffraction peak is lacking due to saturation of the APD.
% 		}
% 	\label{fig:S9}
% \end{figure}

\end{document}